\documentclass[11pt]{article}

\usepackage[normalem]{ulem}
\usepackage{amsmath, amssymb,amsthm,bm,bbm,fullpage}
\usepackage{mathrsfs}
\usepackage{cite}
\usepackage{colortbl}
\usepackage[all]{xy}
\usepackage{epsfig}
\usepackage{subfigure}
\usepackage{graphics}
\usepackage{multirow}
\usepackage{lineno}
\usepackage{setspace}
\usepackage{caption}

\usepackage{graphicx}

\theoremstyle{remark} 
	
\begin{document}

\begin{centering}
{\huge
\textbf{Inequality, Identity, and Partisanship: How redistribution can stem the tide of mass polarization}
}
\bigskip
\\
Alexander J. Stewart$^{1,*}$, Joshua B. Plotkin$^{2}$ and Nolan McCarty$^{3,*}$
\\
\bigskip
\end{centering}
\begin{flushleft}
{\footnotesize
$^1$ School of Mathematics and Statistics, University of St Andrews, St Andrews, KY16 9SS, United Kingdom
\\
$^2$ Department of Biology, University of Pennsylvania, Philadelphia, PA, USA
\\
$^3$ School of Public and International Affairs, Princeton University, Princeton, NJ, USA
\\
$^*$ E-mail: ajs50@st-andrews.ac.uk; nmccarty@princeton.edu
}
\end{flushleft}




\noindent \textbf{The form of political polarization where citizens develop strongly negative attitudes towards out-party policies and members has become increasingly prominent across many democracies. Economic hardship and social inequality, as well as inter-group and racial conflict, have been identified as important contributing factors to this phenomenon known as ``affective polarization.''  Such partisan animosities are exacerbated when these interests and identities become aligned with existing party cleavages. In this paper we use a model of cultural evolution to study how these forces combine to generate and maintain affective political polarization. 
We show that economic events can drive both affective polarization and sorting of group identities along party lines, which in turn can magnify the effects of underlying inequality between those groups. But on a more optimistic note, we show that sufficiently high levels of wealth redistribution through the provision of public goods can counteract this feedback and limit the rise of polarization. We test some of our key theoretical predictions using survey data on inter-group polarization, sorting of racial groups and affective polarization in the United States over the past 50 years.}
\\
\\
\noindent The political polarization of ordinary citizens is increasingly a concern throughout the world, as populist movements challenge mainstream parties in efforts to disrupt established institutions and democratic norms \cite{mudde2017populism}. Such trends have been especially manifest in the United States, where they have culminated in political violence such as of the Unite the Right rally at Charlottesville and the storming of the Capitol during the certification of the 2020 presidential election \cite{Ahler:2018} (see also \cite{WashPo21}).

There has been extensive debate about the nature and causes of mass polarization. The earliest work, focused on the distributions of voter policy preferences, cast considerable doubt as to whether mass polarization was an important phenomenon at all.  That work continues to show that the public's attitudes on policy issues have remained stable and centrist over many decades \cite{Fiorina:2005} (but see \cite{Abramowitz:2010}. This debate is reviewed in \cite{McCarty:2019}). 

However, two other important facets of mass polarization have been rising.  The first is the process of partisan sorting, where the policy preferences and group identities of a voter better align with her partisan attachments \cite{Levendusky:2009,Mason2015,Mason2018}. The second is affective polarization, whereby individuals develop negative attitudes and behaviors towards members of the opposing party \cite{Iyengar2012,Mason2016}.  

Sorting and affective polarization appear to be strongly related to increasing inter-group conflict. The growth of inter-group antagonism has been shown to have multiple contributing factors, including economic adversity, racial animus and a range of other socio-economic factors \cite{Schaffner:2016,Luttig:2017,Sides:2017,Inglehart:2016,Tesler16,Arnorsson:2016,Kolko16,Mitrea20}. Recent work focusing on the cultural evolution of polarization along identity group lines \cite{Stewarteabd4201} has shown that a rise in economic adversity or inequality can cause polarized behavioral strategies to take hold and become entrenched in a population, even when the adverse conditions that stimulated it are reversed \cite{BoleslawWS}.

Despite the important link between partisan sorting and inter-group conflict, there have been few analytical efforts to examine the joint dynamics of these processes.  So in this paper, we generalize the framework of Stewart et al. \cite{Stewarteabd4201} to study the cultural evolution of group polarization and party sorting. In this model, out-group economic interactions are assumed to be more beneficial but more risky than in-group interactions \cite{ruef2002strong,woolley2010evidence,Martinez11,Carruthers96}; and adverse economic environments are assumed to favor risk aversion. We show that when agents attend to both group and partisan identities in choosing interaction partners this stimulates both the evolution of behavioral strategies that polarize along party lines as well as the sorting of group identities along party lines. These behaviors evolve in response to shifts in the economic environment and underlying inequality.

Efforts to mitigate risk aversion in disadvantaged groups via wealth redistribution, in the form of public goods, has the potential to counteract feedback loops that induce polarization. 
And yet, we show that low levels of redistribution can actually magnify underlying inequality and entrench polarization. 
But more optimistically, we also find that sufficiently high levels of redistribution can indeed reduce the impact of inequality and even prevent the emergence of polarization. 

\section*{Results}

To study the cultural evolution of mass political polarization, we generalize a previous model developed to study inter-group polarization and economic interactions \cite{Stewarteabd4201}.

In this model we assume that a large population of individuals is comprised of two distinct identity groups. These identities are assumed fixed, and thus correspond to a fixed feature of identity such as race, religious heritage, or socio-economic background.  Although such identities are fixed in the model, the salience of the identity and therefore its impact on behavior varies.

\begin{figure}
\centering
\includegraphics[width=0.5\linewidth]{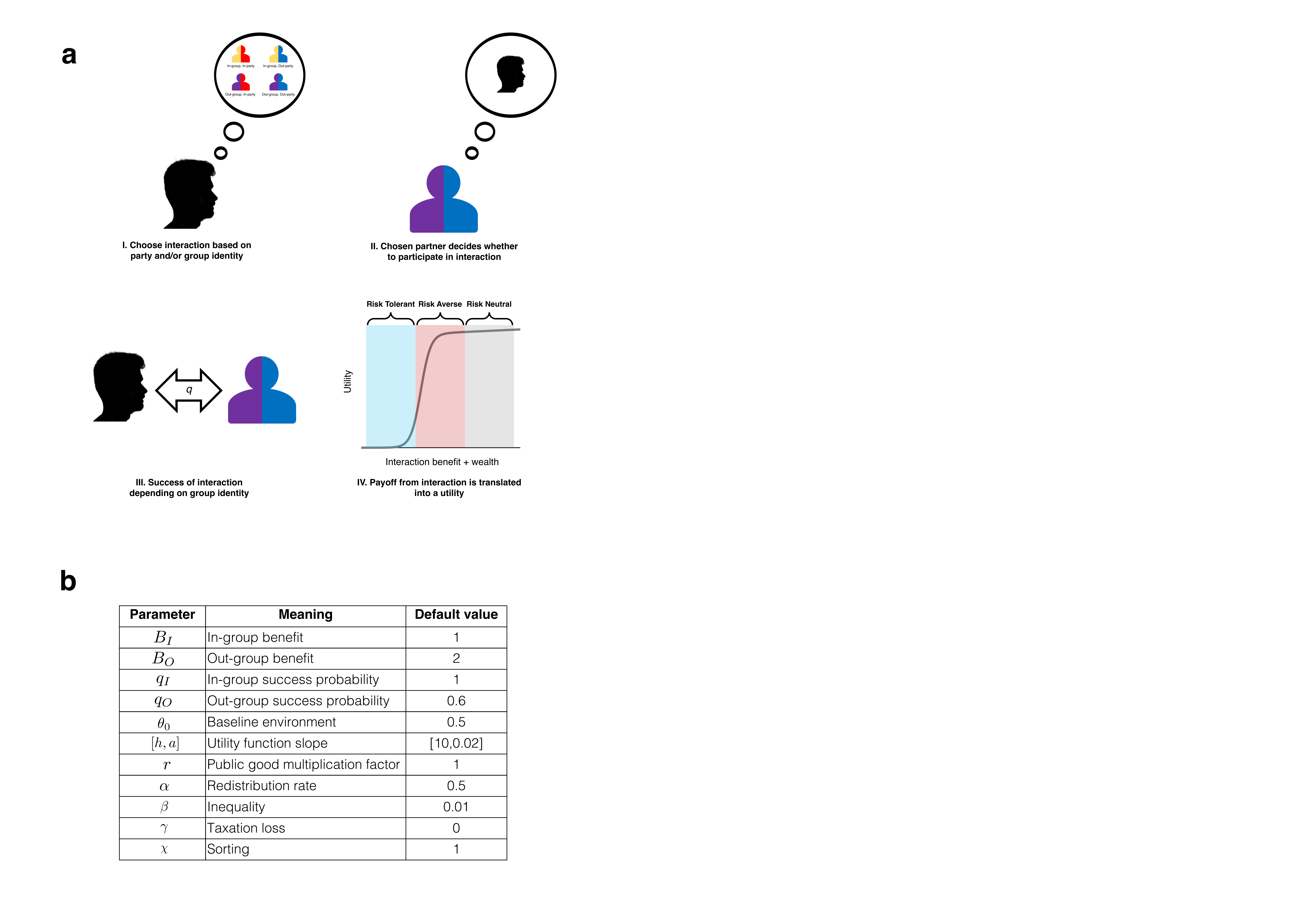}
\caption{\small \textbf{Model of social interaction and identity}. A focal individual (black) engages in beneficial economic interactions. a) I. He first chooses a target for interaction (see Table 1). In general this decision may be based on both party and group identity. II. The chosen target may then agree to engage or not, based on the identity of the focal individual. III. If the pair interacts, the interaction is successful with probability $q_I$, if they share the same identity group, or $q_O$ if they belong to different identity groups. IV. The benefit of a successful interaction is translated into a level of utility that depends on the underlying economic environment (denoted ``wealth'') experienced by the focal individual. Depending on that environment, the utility function he experiences may be risk-neutral (grey region), risk-averse (red region) or risk-tolerant (blue region), as described in Eq.~2. b) The table shows default parameters for our analysis, although we vary these systematically in the SI and show that our results are robust to parameter choice.}
\end{figure}

We assume that members of the population choose to interact with one another, using a one-dimensional strategy $p$, that describes the probability of choosing an in-group member for an economic interaction, whereas the probability of choosing an out-group member is $1-p$ (Figure 1 and Table 1).

An in-group interaction has success probability $q_I$, while an out-group interaction has success probability $q_O$, such that $q_I>q_O$ -- i.e out-group interactions are more risky than in-group interactions. Successful in-group interactions generate benefit $B_I$, whereas out-group interactions generate benefit $B_O$, such that the expected benefit of out-group interactions exceeds that of in-group interactions -- i.e. $q_IB_I<q_OB_O$.

Finally we assume that the state of the underlying economic environment, $\theta$, determines the risk profile experienced by individuals as the benefits of their social interactions are translated into utility. The expected utility for a player $i$ is given by
\begin{eqnarray}
\nonumber w_{i}=p_iq_IF(B_I+\theta)+p_i(1-q_I)F(\theta)+\\
 (1-p_i)(1-\bar{p}_O)q_OF(B_O+\theta)+
(1-p_i)(\bar{p}_Oq_O+(1-q_O))F(\theta)
\end{eqnarray}
\\
where $\bar{p}_O$ is the average strategy among out-group members. We have assumed that out-group interactions are only possible if both players are willing to interact with out-group members, whereas in-group interactions are always available (see Table 1 below and Methods). The function $F$ defines an individual's utility as a function of material payoff $x$ and has the form
\begin{equation}
F(x)=\frac{\exp[hx]}{1+\exp[hx]}(1+a x).
\end{equation}
\\
Here $h$ controls the steepness of the non-linear sigmoid component of the curve and $a$ controls the gradient of the linear component of the curve. This modified `S' shaped utility function allows us to capture changes to risk aversion experienced by individuals as a function of the underlying economic environment, $\theta$. Assuming $a\ll1$, 
the utility function $F$ is maximally concave (risk averse) when $hx\approx\ln\left[\frac{\sqrt{3}+1}{\sqrt{3}-1}\right]$, and is maximally convex (risk-tolerant) when $hx\approx\ln\left[\frac{\sqrt{3}-1}{\sqrt{3}+1}\right]$, and it becomes linear (risk neutral) when $x\gg0$.

Intuitively, our utility function means there is risk-aversion when the underlying economic environment parameter $\theta$ is small but positive. In this regime, which may be thought of as analogous to a risk profile of an individual close to insolvency, failures of economic interactions result in very sharp declines in utility -- and so in-group interactions are preferable to the more risky out-group interactions. But when the underlying economic environment is very good ($\theta\gg0$), risk aversion declines and out-group interactions, which have greater expected returns, are preferable. Finally when the underlying environment is so bad ($\theta<0$) that a successful economic interaction produces a sharp increase in utility, then risky out-group interactions become strongly preferred.

Under this model of inter-group interaction, it has already been shown  \cite{Stewarteabd4201} that both high polarization ($p=1$) and low polarization ($p=0$) are stable outcomes when the economic environment is strong; but only high-polarization is stable as risk aversion increases. As a result, populations tend to become polarized when the underlying economic environment exogenously declines, and they remain polarized even if the economic environment subsequently improves.

We now generalize this framework to study mass political polarization, in which individuals have a fixed group identity and a sticky, but more malleable, partisan identity. We study how polarization along party lines can emerge as a consequence of risk aversion, as well as the extent to which group identities sort along party lines. We further allow for feedback between individuals' economic interactions and the overall state of the economic environment, so that the environmental dynamic is not exogenous, but rather coupled to partisan identification and individual decisions about economic interactions. This coupling leads to a runaway process that accelerates the rise of polarization and also exacerbates economic inequality through its impact on inter-group interactions.
\\
\\
\noindent\textbf{Model of party identity and social decisions:} In order to generalize the model outlined above, and to capture the dynamics of mass political polarization, we assume that the population is composed of two identity groups and two political parties, such that each individual has both a group identity and a party identity. In general we assume an individual's  partisan identity can change, while their group identity is fixed. The risks and benefits of economic interactions between individuals vary by group identity, but we assume that they are independent of party identity.

We now consider two additional decision processes beyond that described above. In these versions, an individual's strategy $p$ depends both on party and group identity. First we describe a case in which the decision to interact with another player depends on their party identity alone, and second we consider a case in which both group identity and party identity are salient to interaction choices.

Table 1 summarizes the parameters and interaction probabilities in all three cases. A detailed description of the mathematical model is given in the Methods section below, and further details of its analysis can be found in the SI.

\begin{figure}[th!] \centering \includegraphics[scale=0.1]{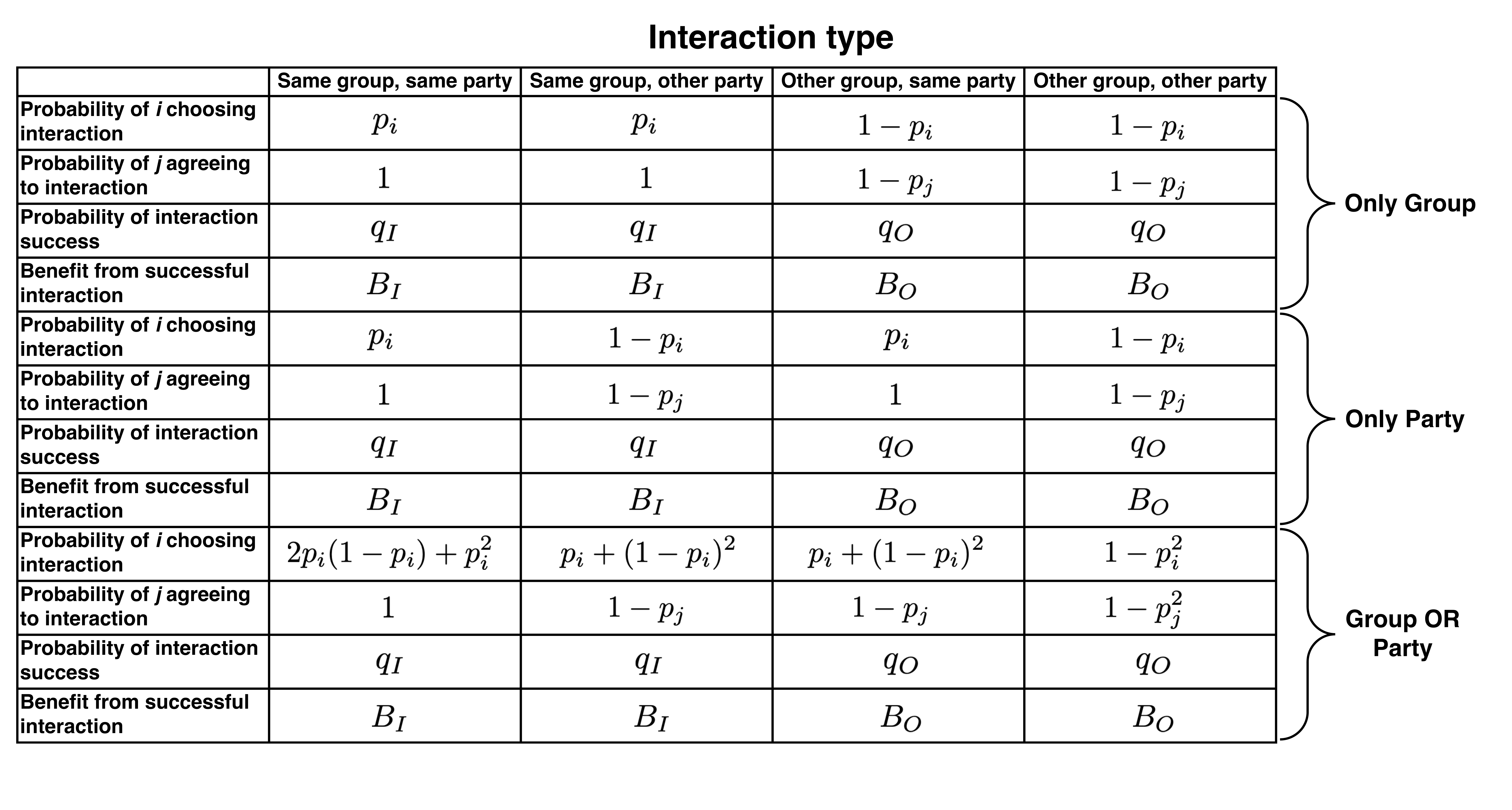}
\caption*{\small Table 1: \textbf{Summary of decision processes}. We consider three decision processes for a focal individual $i$ choosing an economic interaction. For each decision process, there is a probability of interaction given the identity of a potential target, and a probability of a target $j$ consenting to interaction. If an interaction takes place, the probability of success, and the benefit generated, depend on the identity groups of the pair.}
\end{figure}

The key differences between the various decision processes summarized in Table 1 involve  different probabilities that a player chooses a particular type of interaction, and different probabilities of that interaction being accepted by the other agent. When group identity alone is salient to choice of interaction partners, then out-group members may reject an interaction. When only party identity is salient, then out-party members may reject an interaction. When both group identity and party identity are salient, either an out-group or an out-party member may reject an interaction.  

We show that for all three decision processes the dynamics of cultural evolution lead to bi-stability when the underlying economic environment makes individuals risk-neutral or risk-tolerant, with both a high-polarization and a low-polarization equilibrium as stable outcomes, and that this bistability is robust to the choice of parameters (see SI). However, if the environment becomes sufficiently risk-averse, only the high-polarization equilibrium is stable. Thus a population faced with a sufficiently risk-averse environment moves towards a state of high polarization and remains there, even if the underlying environment subsequently improves and risk aversion declines \cite{Stewarteabd4201}. We explore the consequences of these dynamics for sorting of identity groups along party lines, and in the presence of redistribution via public goods. 
\\
\\
\textbf{Sorting:} In general, political parties and identity groups may be different in size. However, we make the simplifying assumption that both groups and parties are equal in size. If we denote the proportion of group 1 in party 1 as $x_1$ and the proportion of group 2 in party 2 as $x_2$, the assumption of equal sized groups and parties and groups means $x_1=x_2=x$. Under this assumption we can define the degree of sorting of identity groups along partisan lines via the simple expression

\begin{equation}
\chi=2x-1
\end{equation}
\\
such that $\chi=0$ corresponds to identity groups distributed equally among the parties, while $\chi=1$ corresponds to party 1 perfectly aligned with group 1 and $\chi=-1$ corresponds to party 2 perfectly aligned with group 1.
\\
\\
\textbf{Inequality and redistribution:} In our model, successful economic interactions not only benefit the pair of interacting individuals, but they also generate a contribution to a public good that benefits the population at large. To capture this public goods provision we assume that the current economic environment $\theta$ is a linear function of the benefits generated by successful interactions:

\begin{equation}
\theta=\alpha(1-\gamma\alpha)r\bar{B}-\theta_0
\end{equation}
\\
where $\alpha$ is the rate of wealth redistribution, $\gamma$ captures the deadweight loss due to taxation, $r$ is the benefit multiplication factor of the public good, and $\theta_0$ is the baseline economic environment when no economic interactions occur. Here $\bar{B}$ denotes the average benefit from economic interactions across the population. The ``after tax'' payoff received by an individual who generates benefit $B$ from economic interactions is thus $(1-\alpha) B+\theta$. This model is motivated by political economic models of linear taxation \cite{Meltzer81}.  

\begin{figure}[th!] \centering \includegraphics[scale=0.25]{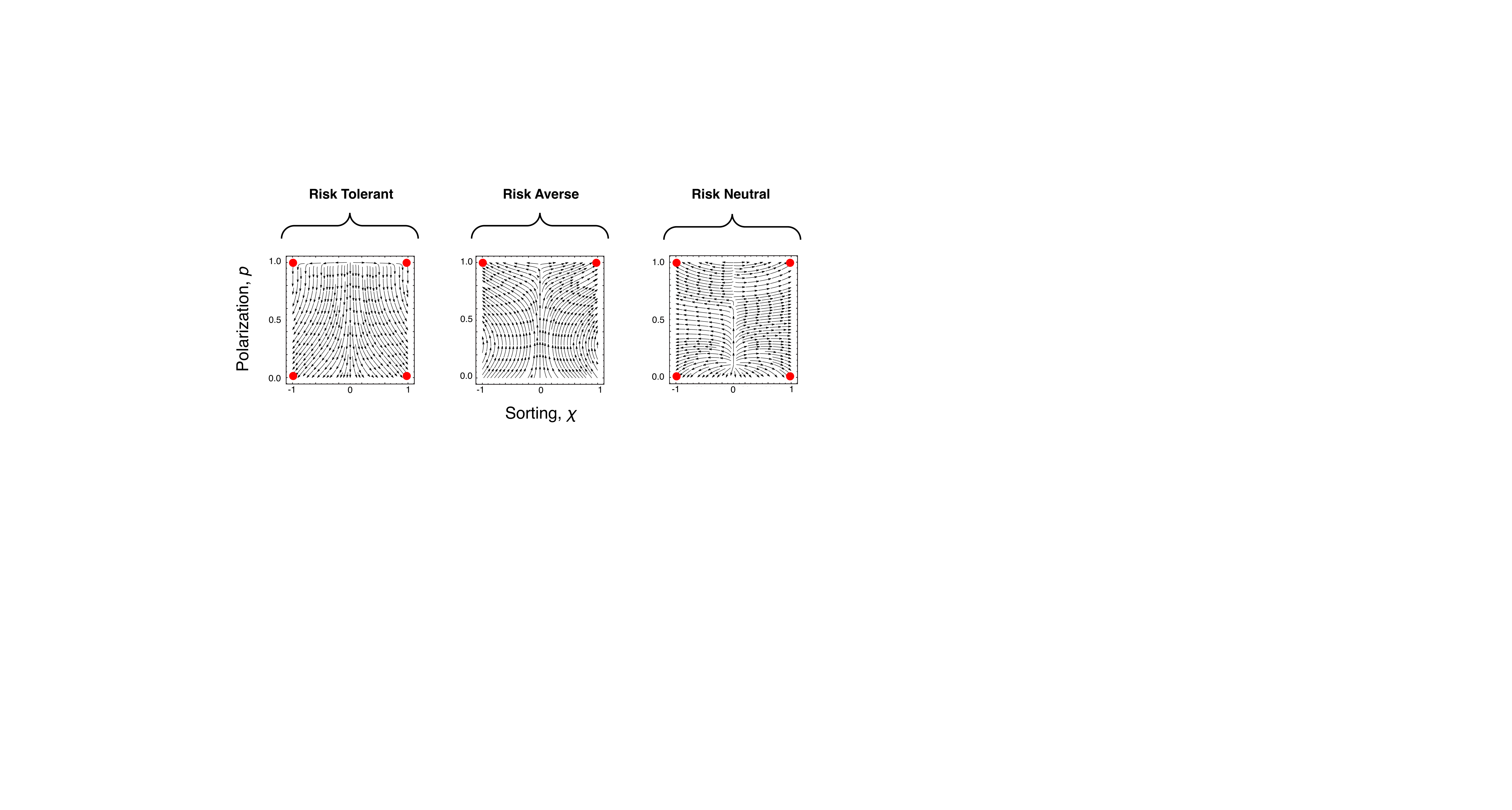}
\caption{\small \textbf{Polarization and sorting}. Phase portraits illustrate the dynamics of polarization $p$ and degree of sorting $\chi$ under our model of economic interactions and party switching, with fixed identity groups. Arrows indicate the average selection gradient experienced by a local mutant against a monomorphic background (see Methods). Red dots indicate stable equilibria. (left) When the decision process for social interactions considers group or party identity, both high and low polarization states are stable, but sorting is always high $|\chi|=1$. (center) However when the environment is risk averse, only high polarization and high sorting are stable. (right) And finally when the environment is risk neutral, the system returns to bistable polarization with high sorting. These plots show dynamics for $B_I=1$, $B_O=2$, $q_I=1.0$, $q_O=0.6$, $h=10$ and $a=0.02$. The phase portraits here show the selection gradient experienced by a monomorphic population in which parties and groups are of equal size. Dynamics under alternate decision processes (Table 1) and different parameter choices are shown in the SI.}
\end{figure}

Redistribution of public goods is particularly important in the presence of pre-existing wealth inequality. A large body of empirical and theoretical work has demonstrated that inequality and polarization correlate and are likely causally linked \cite{McCarty16}. To capture the effects of pre-existing inequality we assume that one identity group receives benefits from social interactions scaled by a factor $2\beta B$ while the other group receives $2(1-\beta)B$. Thus when $\beta =0.5$ there is no baseline inequality, whereas when $\beta=0.01$ the wealthier group receives around $100$ times greater benefits than the poorer group, per interaction.  
\\
\\
\textbf{Joint dynamics of sorting and polarized attitudes:} We first consider the interdependence of sorting and polarization, keeping party size fixed such that a small change in $x$ can be thought of as two members of different identity groups and parties swapping parties. We explore this interdependence for decision processes that account for group and party, 
and those that account for party alone.

We find that for both types of decision processes, low polarization favors high sorting -- so that people change parties until parties are aligned with identity group. This is in contrast to a decision process that takes account only of group identity, where sorting has no effect on utility and therefore does not tend to evolve (see SI). Under the mixed scenario, in which interaction strategies attend to both group and party identity, high sorting evolves in all environments (Figure 2). And so the model predicts, in general, that a shift from individuals paying attention to only group identity, to individuals also paying attention to party identity, will lead to sorting.

For interaction strategies that account only for party identity, however, high polarization favors low sorting (Figure S2) when the environment does not induce risk aversion. This arises because, when players focus only on party identity and only interact with their own party, expected payoffs can be maximized by making the parties well mixed with respect to identity group, without any risk of failed interactions due to out-group members refusing to participate in an interaction (Table 1). However when group identity is visible and salient for interactions, this is not possible.
\\
\\
\noindent\textbf{Race, party and sorting:}
Our model predicts that individuals who take into account both group and party identity when making economic decisions will tend to evolve to a state in which group and party identity align. To test this theoretical prediction we looked at the salience of racial identity, party identity, and sorting of people identifying as white in US presidential elections between 1964 and 2016. Using American National Election Survey (ANES) data we calculated the affective polarization and in-group favorability towards white versus non-white people, as well as the degree of sorting, measured by the variance in party preference explained by racial identity (Figure 3 and SI) \cite{ANES}. 

\begin{figure}[th!] \centering \includegraphics[scale=0.25]{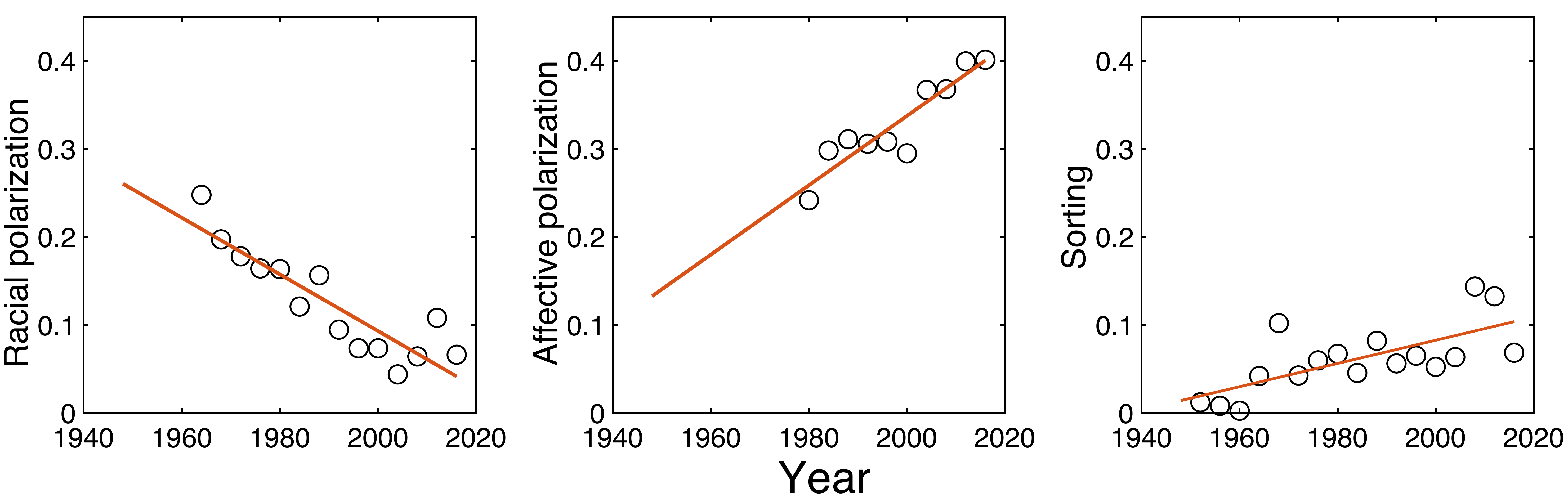}
\caption{\small \textbf{Increasing salience of party identity in the United States.} We looked of in-group favorability (left) and affective polarization (center) among white respondents in ANES data at each presidential election over past decades. While in-group favorability towards white versus non-white people declines over time ($p<0.01,\ \ t=6.9$), affective polarization increases ($p<0.01,\ \ t=6.7$) indicating a relative decline in the salience of racial identity and an increase in party identity. At the same time sorting of racial groups along party lines has increased (right, $p<0.01,\ \ t=3.7$), as measured by the variance in party preference explained by white vs non-white identity.}
\end{figure}

We find that the salience of racial identity (measured by the in-group favorability among white respondents) has declined over time, while the salience of party has increased (measured by affective polarization among white respondents) as sorting has increased. This suggests a shift in which individuals pay relatively greater attention to party identity over time, and also become more sorted with respect to racial identity.  This pattern is consistent the predictions of our model (Figure 2) -- namely, that as attention is increasingly paid to party, this will induce sorting of group identities along party lines.
\\
\\
\textbf{Redistribution, inequality and polarization:} According to our model increased polarization arises as a result of risk aversion in a poor economic environment. In general, however, different identity groups may experience different economic environments. In particular, when there is inequality such that some groups possess less wealth than others, they are more likely to be risk averse and thus become polarized. Such inequality can lead to the evolution of polarization in the whole population \cite{Stewarteabd4201}. However, redistribution via public goods can reduce inequality, and might improve the overall economic environment. 

We use Monte Carlo simulations to explore the impact of such redistribution on the dynamics of mass polarization in the presence of inequality. In particular we focus on situations in which the range of $\theta$ (Eq.~4) encompasses different economic environments, ranging from risk-neutral through risk-averse and risk-tolerant.

\begin{figure}[th!] \centering \includegraphics[scale=0.2]{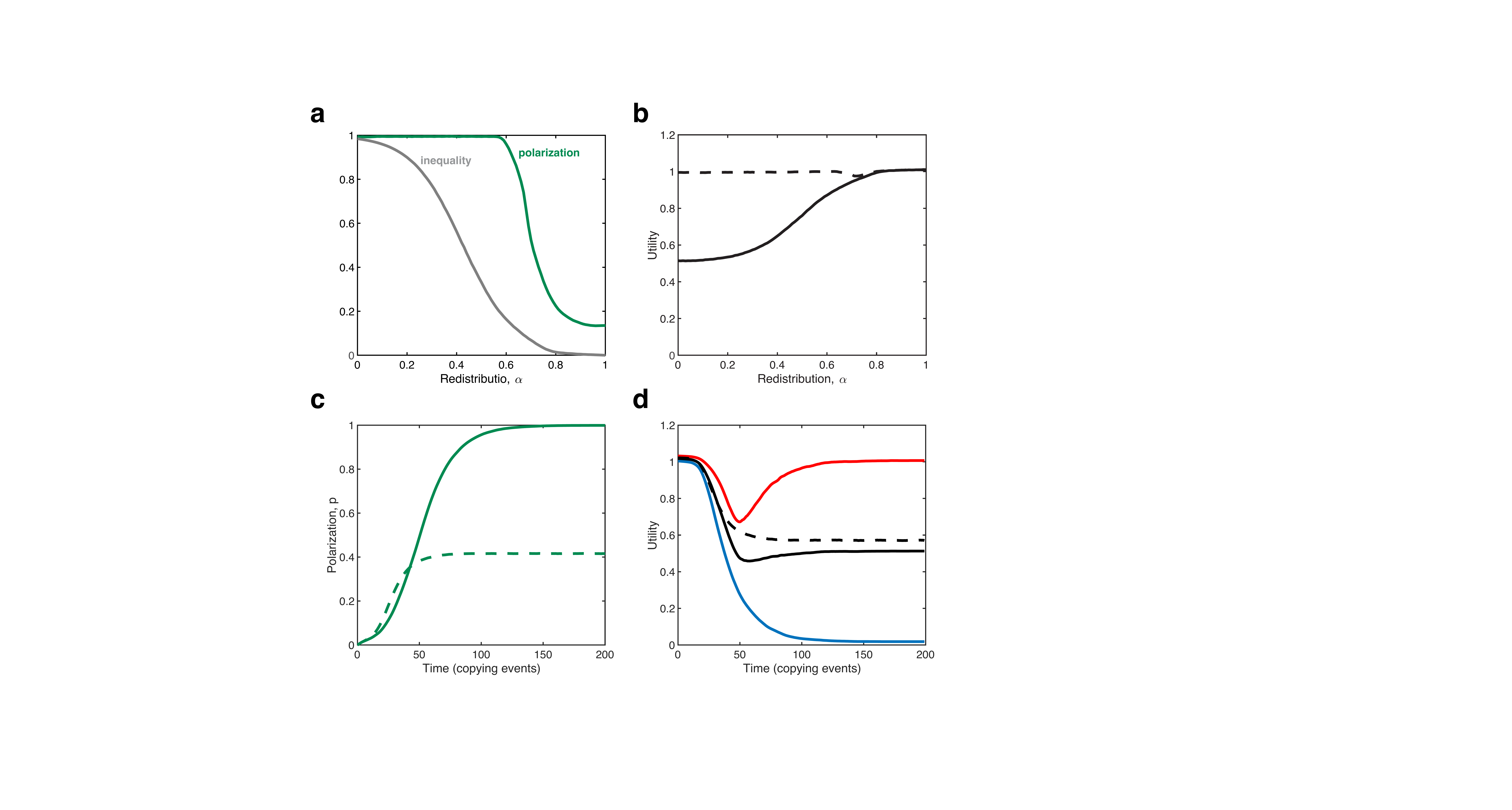}
\caption{\small \textbf{Redistribution and inequality.} Ensemble mean equilibria (panels a and b) and time trajectories (panels c and d)  for a population initialized in a low polarization state, from individual-based simulations in the presence of wealth redistribution (Eq.~4). We show results in the case of no underlying economic inequality, $\beta=0.5$ (dashed lines), as well the case of high underlying inequality, $\beta=0.01$ (solid lines). Results shown here arise from a decision process that attends to group or party identity, and sorting is fixed exogenously at $\chi=1$. (a) When public goods are not multiplicative ($r=1$ and $\theta_0=0.5$), and redistribution is absent ($\alpha=0$) overall inequality (gray line, measured as the relative difference in utility -- see SI) and polarization (green line) are high. With increasing rates of redistribution, first overall inequality and then polarization decline to zero. b) Increasing redistribution increases overall utility towards the level achieved when underlying inequality is absent. c) When public goods increase overall utility ($r=10$ and $\theta_0=5.4$), redistribution can act to magnify both polarization and inequality over time, compared to a fixed environment without feedback via redistribution. 
d) Overall inequality initially has a relatively small impact on population mean utility in a population initialized in a low-polarization state (black lines), but redistribution is seen to magnify the effects of underlying inequality, with the richer group (red line) experiencing a transient decline in utility as polarization evolves before returning to a value close to maximum, while the poorer group (blue line) suffer an irreversible decline towards utility close to zero. Plots show ensemble mean values across $10^4$ replicate simulations, for groups of 1000 individuals each. Success probabilities and benefits are fixed at $B_I=1$, $B_O=2$, $q_I=1.0$, $q_O=0.6$ with $h=10$ and $a=0.02$, while $\gamma=0$. Evolution occurs via the copying process (see methods) with selection strength $\sigma=10$, mutation rate $\mu=10^{-3}$ and mutation size $\Delta=0.01$.}
\end{figure}

Figure 4 shows the effect of redistribution and inequality on the dynamics of polarization. We see that sufficient redistribution can reduce both inequality and polarization, although a high degree of redistribution is required to prevent polarization. This effect holds when public goods are purely redistributive ($r=1$ in Eq. 4), and when public goods increase the overall wealth of the population ($r>1$) (Figure S10). When taxation produces a deadweight loss ($\gamma>0$, Eq. 4), it becomes harder to reduce polarization via redistribution (Figure S3).

 Although sufficient redistribution can reduce inequality and polarization, it is also important to note that the effect of feedback between individual economic interactions and the overall economic environment that arises as a result of redistribution can facilitate the evolution of polarization compared to a high quality stable environment (i.e fixed $\theta$ -- Figure 4c) in which polarization does not evolve. Thus introducing feedback between individual interactions and the environment through low or intermediate levels of redistribution can make things worse, by both failing to reduce inequality and facilitating the evolution of polarization (Figure 4 c-d).



We exogenously varied the amount of sorting, $\chi$. Sorting tends to increase polarization, but it can have complex effects on levels of inequality and population average utility. This is because intermediate levels of polarization tend to result in lower levels of utility. Where reducing sorting can also reduce polarization to low levels, it has a beneficial effect in reducing inequality and increasing population average utility (see SI). 
\\
\\
\textbf{Inequality reduces average utility:} We also explored the impact of inequality on population average utility. In general, underlying inequality resulting from lower income from economic interactions, tends to reduce the population average utility compared to the case where average income remains the same, but there is no inequality (Figure 4 and Figure S9-S10). This is because the poorer group tends to experience the risk averse environment in which failed interactions produce a sharp decline in utility.
\\
\\
\textbf{Recovery from polarization:} For a wide range of parameter values the system is bistable, with both high- and low-polarization equilibria maintained unless the environment is risk averse (see SI). Up until now we have focused on the conditions under which a population will evolve from a low- to a high-polarization state -- i.e. the conditions under which the low-polarization equilibrium is lost. However, recovering low polarization once high polarization has evolved requires a switch from one equilibrium to another. In practice this may occur as the result of an environmental shock (see SI) or as a result of coordinated action in which a sufficient number of individuals simultaneously adopt a low-polarization strategy to move the population to a state that is then attracted to a low-polarization equilibrium. The threshold frequency of individuals required to achieve this transitions is determined by the size of basin of attraction for the high-polarization equilibrium.

We calculated the frequency required for escape -- the proportion of the population that must simultaneously adopt a low-polarization behavior to escape the high-polarization equilibrium (Figure 5). We find that, for many environments, escape is only possible once the economic environment is sufficiently advantageous, at which point escape from polarization can become a realistic prospect. Thus, addressing polarization requires first improving the economic environment and then engaging in sufficient coordinated action to adopt low polarization behavior, followed by the spread of low-polarization behavior by social contagion.
\\
\\
\begin{figure}[th!] \centering \includegraphics[scale=0.15]{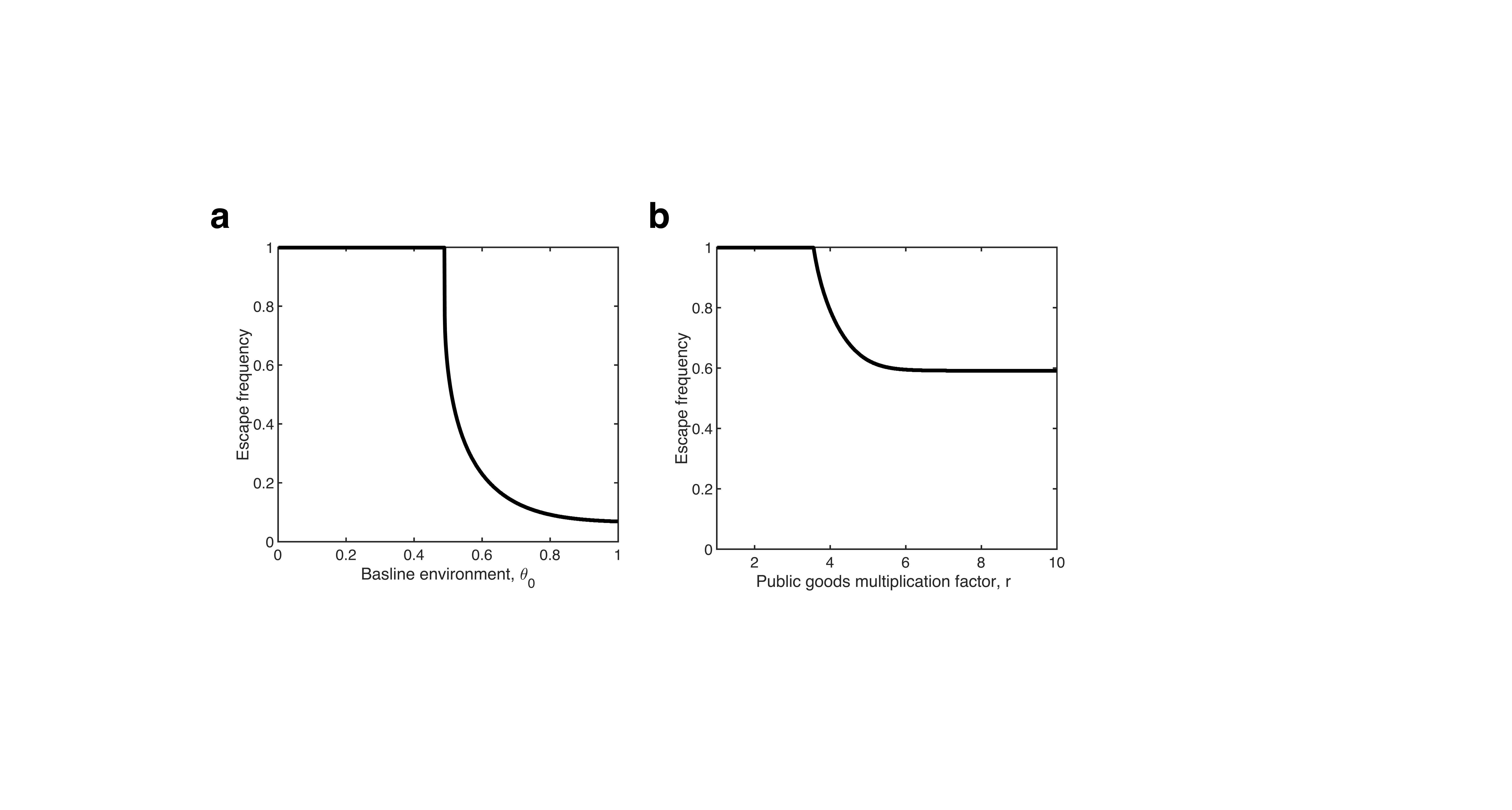}
\caption{\small \textbf{Recovering from polarization.} We numerically calculated the size of the basin of attraction for the high-polarization state, which determines the escape frequency -- that is, the proportion of the population that must simultaneously adopt a low-polarization behavior to escape the high-polarization equilibrium. a) As the baseline environment, $\theta_0$, improves, it becomes increasingly easy to reverse polarization through the coordinated behavior of small frequency of low-polarization individuals. b) Similarly, increasing the value of public goods, $r$, makes it easier to reverse polarization. Here other parameters are set to $B_I=1$, $B_O=2$, $q_I=1.0$, $q_O=0.6$ with $h=10$ and $a=0.02$, while $\gamma=0$, $\beta=0.5$, $\theta_0=0.5$ and $r=1$ unless otherwise stated.}
\end{figure}

\section*{Discussion}


Our model provides a framework for connecting the effects of inter-group animus, economic adversity and mass political polarization through the lens of cultural evolution \cite{Boyd85,Cavalli81}. We focus on polarization expressed through loss of positive social interactions with members of an out-party \cite{Iyengar2012,Mason2016}, and sorting of identity groups along party lines \cite{Levendusky:2009,Mason2015,Mason2018}. We show that attending to party identity when deciding who to interact with is sufficient to translate inter-group polarization stimulated by adverse economic conditions into political polarization between members of opposing parties (Figure 2). We further show that if party identity is able to evolve alongside behavioral strategies, this can also lead to sorting of identity groups along party lines (Figure 3). We then show that feedback between individuals' economic interactions and the overall economic environment can lead to increased polarization and amplify the effects of underlying inequality between groups (Figure 4). These effects are magnified when identity groups are sorted along party lines, but can be mitigated or prevented entirely if redistribution of wealth via public goods is put in place to combat inequality (Figure 4-5).

In performing simulations (Figure 4) we used fixed levels of sorting, on the basis that changes to party identity can be assumed slow compared to changes in interaction strategy. However if the evolution of sorting occurs rapidly, situations may arise in which members of different parties experience divergent selection pressures (see SI), raising the possibility that different levels of polarization may arise in some subsets of the population and not others. Exploring this possibility provides a natural direction for further work. 

Our work focuses on affective polarization and sorting with respect to identity groups among the electorate. However attempts to prevent and reverse polarization must also take account of the mechanisms that enable elite \cite{FeldmanWS,LeonardWS} and ideological polarization \cite{AxelrodWS,LeonardWS}, and must account for the role of factors such as geography \cite{ChuWS} and population and social network structure \cite{Stewart19,SantosWS,TokitaWS} in producing mass polarization, in addition to the inter-group and economic factors studied here. We must also remain alert to the circumstances under which polarization can provide benefits \cite{KawakatsuWS,VasconcelosWS} (e.g.~Figure S9 in which increased sorting can increase polarization but reduce inequality).

The impact of underlying inequality on the evolution of polarization, and the amplification of the effects of inequality via economic feedback, illustrate the need to think carefully about mass political polarization in the context of inter-group conflict and the economic environment \cite{McCarty16, Ahler:2018}. This is particularly true when assessing ways to prevent or reverse mass polarization. The success of redistribution at stemming the tide of polarization in our model is striking, and it suggests a possible path for preventing such attitudes from taking hold in future. We emphasize though that this strategy is only possible if implemented in a population that is not already polarized, in an environment that supports low polarization. Once polarization sets in, it typically remains stable under individual-level evolutionary dynamics, even when the economic environment improves or inequality is reversed. The only remedy for reversing a polarized state, under our analysis, requires either a shock (Figure S11) or a sufficiently good economic environment coupled with collective action by a portion of the population who change strategies simultaneously.

\section*{Methods}

In this section we describe the decision process, calculation of utility and selection gradient, and the copying process used in simulations. Further analysis of the model can be found in the SI.
\\
\\
\textbf{Measure of inequality:}
Throughout we adopt a simple measure of inequality: the difference in relative utility between the two groups i.e. $\frac{w_{HIGH}-w_{LOW}}{w_{HIGH}+w_{LOW}}$ where $w_{HIGH}$ is the average utility of the richer group and $w_{LOW}$ the utility of the poorer group
\\
\\
\textbf{Decision process:} Table 1 gives the probability for a focal player $i$ choosing to interact with a given player $j$ based on the identity of $j$ and the decision process adopted by $i$. In order to calculate the utility of $i$ given a decision process, we must calculate  the probability distribution for the next interaction $i$ participates in, conditional on an interaction occurring. That is, we must weight the probability of interactions given in Table 1 by the number of individuals in each group, and normalize the distribution. This corresponds to a process in which the focal player randomly draws an individual from the population and then decides to pursue an interaction with that individual based on the probabilities given in Table 1. These normalized distributions are given below for the decision process that takes account of only party identity, and for the decision process that takes account group or party identity. Note that if the decision process takes account of group identity only, no normalization is required, since the degree of sorting does not impact the probability of interaction.   
\\
\\
\textbf{Only party identity:} Under this decision process the probability of an individual $i$ belonging to group 1 and party 1 choosing to interact with an individual with identity $kl$ is $\pi_{kl}(x)$ where $k$ indexes the group identity $I$ or $O$ and $l$ indexes the party identity and $x$ is the frequency of individuals from group 1 in party 1 (and, by symmetry, the number of individuals from group 2 in party 2). We then have

\begin{eqnarray}
\nonumber\pi_{II}(x)=\frac{p_ix}{p_ix+(1-p_i)(1-x)+p_i(1-x)+(1-p_i)x}\\
\nonumber\pi_{IO}(x)=\frac{(1-p_i)(1-x)}{p_ix+(1-p_i)(1-x)+p_i(1-x)+(1-p_i)x}\\
\nonumber\pi_{OI}(x)=\frac{p_i(1-x)}{p_ix+(1-p_i)(1-x)+p_i(1-x)+(1-p_i)x}\\
\nonumber\pi_{OO}(x)=\frac{(1-p_i)x}{p_ix+(1-p_i)(1-x)+p_i(1-x)+(1-p_i)x}\\
\end{eqnarray}
\\
where $x$ is the proportion of identity group $k$ that also belong to party $k$.
\\
\\
\textbf{Party OR Group Identity:} Under this decision process the probability of an individual $i$ who belongs to group 1 and party 1, choosing to interact with an individual with identity $kl$ is $\phi_{kl}(x)$. We then have

\begin{eqnarray}
\nonumber \phi_{II}(x)=\frac{(2p_i-p_i^2)x}{(2p_i-p_i^2)x+2(p_i+(1-p_i)^2)(1-x)+(1-p_i^2)x}\\
\nonumber \phi_{IO}(x)=
\frac{(p_i+(1-p_i)^2)(1-x)}{(2p_i-p_i^2)x+2(p_i+(1-p_i)^2)(1-x)+(1-p_i^2)x}\\
\nonumber \phi_{OI}(x)=\frac{(p_i+(1-p_i)^2)(1-x)}{(2p_i-p_i^2)x+2(p_i+(1-p_i)^2)(1-x)+(1-p_i^2)x}\\
\nonumber \phi_{OO}(x)=\frac{(1-p_i^2)x}{(2p_i-p_i^2)x+2(p_i+(1-p_i)^2)(1-x)+(1-p_i^2)x}\\
\end{eqnarray}
\\
This decision strategy reflects a situation in which an individual sees someone as a member of their in-group if they share either the same group or the same party identity, and weights both of those dimensions of identity equally. We explore an AND type decision process in the SI.
\\
\\
\textbf{Expected utility:} In order to explore the evolutionary dynamics of polarization, we calculate the expected utility of a mutant strategy $p_i$, which deviates by a small amount from the resident strategy $p$ employed by the rest of the population. Using Eq. 5 and Eq. 6 above we can now write down the expected fitness for such a mutant under a given decision process. When players only attend to party identity the utility of such a mutant is

\begin{eqnarray}
\nonumber w_i(x)=[\pi_{II}(x)+\pi_{IO}(x)(1-p)]q_IF(B_I+\theta)+\\
\nonumber[(\pi_{II}(x)+\pi_{IO}(x))(1-q_I)+\pi_{IO}(x)pq_I]F(\theta)+\\
\nonumber[\pi_{OI}(x)+\pi_{OO}(x)(1-p)]q_OF(B_O+\theta)+\\
\nonumber[(\pi_{OI}(x)+\pi_{OO}(x))(1-q_O)+\pi_{OO}(x)pq_O]F(\theta)\\
\end{eqnarray}
\\
whereas the utility of a mutant when players attend to party or group is

\begin{eqnarray}
\nonumber w_i(x)=[\phi_{II}(x)+\phi_{IO}(x)(1-p)]q_IF(B_I+\theta)+\\
\nonumber[(\phi_{II}(x)+\phi_{IO}(x))(1-q_I)+\phi_{IO}(x)pq_I]F(\theta)=\\
\nonumber[\phi_{OI}(x)(1-p)+\phi_{OO}(x)(1-p^2)]q_OF(B_O+\theta)=\\
\nonumber[(\phi_{OI}(x)+\phi_{OO}(x))(1-q_O)+(\phi_{OI}(x)p+\phi_{OO}(x)p^2)q_O]F(\theta)\\
\end{eqnarray}
\\
\textbf{Selection gradient:} We can now calculate the average selection gradient experienced by the mutant $p_i$, which is given by

\begin{equation}
s_p=\frac{\partial [xw_i(x)]}{\partial p_i}\Bigg|_{p_i=p}+\frac{\partial [(1-x)w_i(1-x)]}{\partial p_i}\Bigg|_{p_i=p}
\end{equation}
\\
When the selection gradient is positive the mutant has an advantage over the resident strategy on average. Note however that when $0<|\chi|<1$ different individuals experience different effects from the same mutation. This issue is discussed in more detail in the SI.

We can also calculate the effect of a small change to the degree of sorting in the population by calculating the gradient \cite{Mullon:2016aa,Leimar}

\begin{equation}
s_x=\frac{\partial [xw_i(x)]}{\partial x}+\frac{\partial [(1-x)w_i(1-x)]}{\partial x}
\end{equation}
\\
When $s_x$ is positive, the average effect of an increase in sorting is to increase the average utility of the population. It is Eqs. 5-10 that are used to produce Figures 2-3 (see also SI).
\\
\\
\textbf{Evolutionary simulations:} In order to simulate the evolutionary dynamics of this system we consider a population evolving under a ``copying process'' \cite{Traulsen:2006aa} in which individuals are able to observe the utility of other individuals and compare it to their own.  The dynamics of the model are as follows: An individual $i$ is chosen at random from a population of fixed size $N$. A second individual $j$ is then chosen at random for her to ``observe''. If $i$ has utility $w_i$ and $j$ has utility $w_j$ then $i$ chooses to copy the strategy of $j$ with probability $1/(1+\exp[\sigma(w_j-w_i)])$, where $\sigma$ scales the ``strength of selection'' of the evolutionary process. Note that if $w_j\gg w_i$ the probability of $i$ copying the behavior of $j$ is close to 1, whereas if $w_j\ll w_i$ the probability is close to 0. 

Individual based simulations used to produce Figures 4-5 were performed under the copying process using populations composed of two identity groups of size $N=1000$ individuals, with sorting of groups among parties fixed. Mean trajectories were determined from an ensemble of $10^4$ sample paths. Simulations were run for $100N-200N$ copying events to find equilibria. Mutations were assumed to occur at a rate $1/N$ per copying event, with the target of the mutation chosen randomly from the population. Mutations were assumed to be local such that the target of the mutation had their strategy perturbed by $\Delta=\pm 0.01$ with mutations that increase and decrease $p$ equally likely, and we impose the appropriate boundary conditions to ensure strategies were physical.

\clearpage

\section*{Supporting Information}

Here we present additional numerical analysis and simulations to demonstrate the robustness of our findings to relaxation of model assumptions.

\section*{Decision Processes}

In this section we describe two additional decision processes, beyond those described in Table 1 of the main text.  Table S1 gives event probabilities for an AND logic decision process, in which an individual only regards another as part of their in-group if they belong to the same identity group and the same party:

\begin{figure}[th!] \centering \includegraphics[scale=0.1]{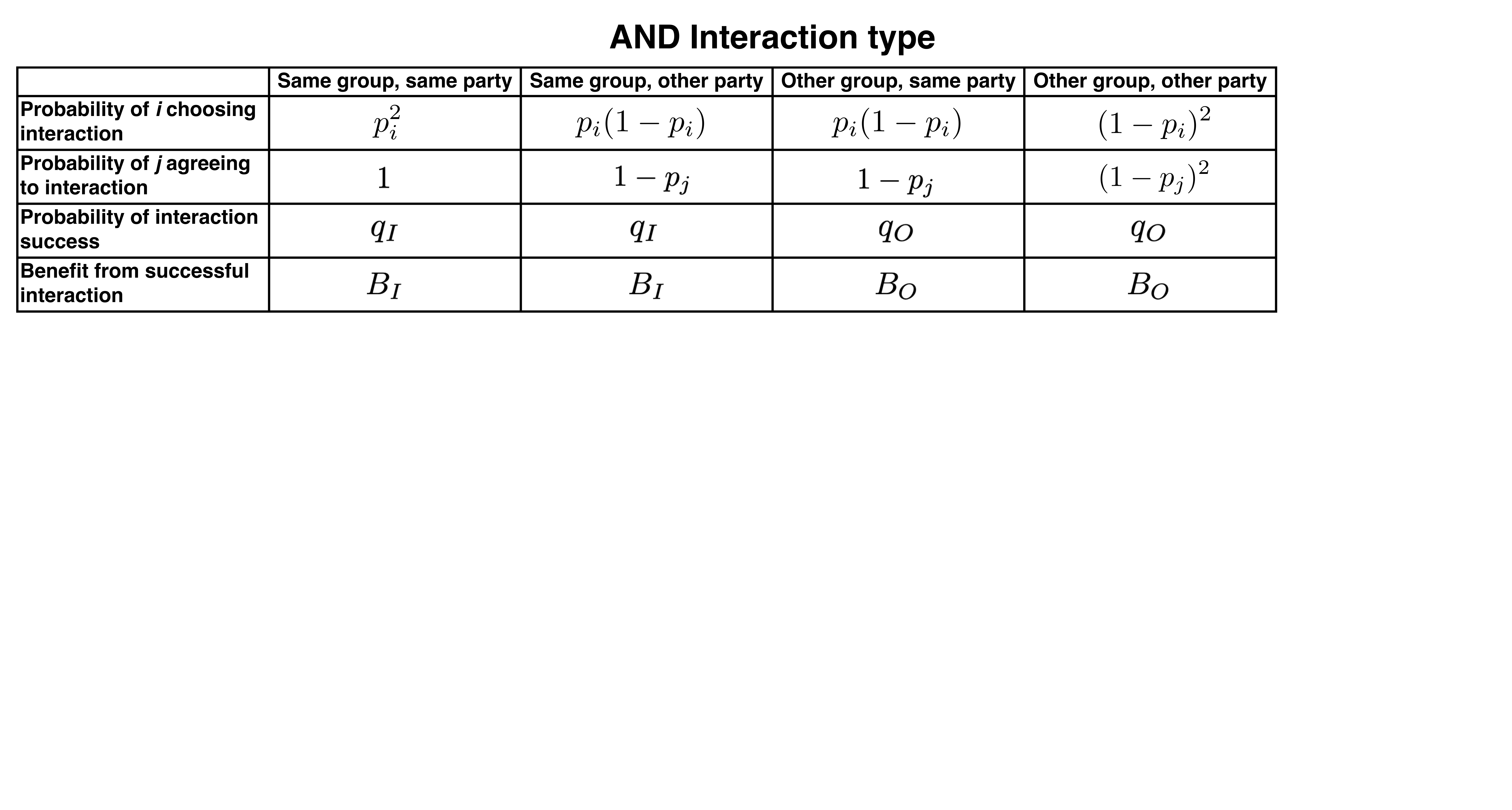}
\caption*{\small Table S1: \textbf{AND decision logic}.}
\end{figure}

Under this decision process the probability of an individual $i$, belonging to group 1 and party 1, choosing to interact with an individual with identity $kl$ is $\lambda_{kl}$. We then have

\begin{eqnarray}
\nonumber \lambda_{II}=\frac{(2p_i-p_i^2)x}{(2p_i-p_i^2)x+(p_i+(1-p_i)^2)(1-x)+(p_i+(1-p_i)^2)(1-x)+(1-p_i^2)x}\\
\nonumber \lambda_{IO}=\frac{(p_i+(1-p_i)^2)(1-x)}{(2p_i-p_i^2)x+(p_i+(1-p_i)^2)(1-x)+(p_i+(1-p_i)^2)(1-x)+(1-p_i^2)x}\\
\nonumber \lambda_{OI}=\frac{(p_i+(1-p_i)^2)(1-x)}{(2p_i-p_i^2)x+(p_i+(1-p_i)^2)(1-x)+(p_i+(1-p_i)^2)(1-x)+(1-p_i^2)x}\\
\nonumber \lambda_{OO}=\frac{(1-p_i^2)x}{(2p_i-p_i^2)x+(p_i+(1-p_i)^2)(1-x)+(p_i+(1-p_i)^2)(1-x)+(1-p_i^2)x}\\
\end{eqnarray}
\\
The utility of a mutant when players attend to party AND group is thus

\begin{eqnarray}
\nonumber w_i(x)=[\lambda_{II}(x)+\lambda_{IO}(x)(1-p)]q_IF(B_I+\theta)+[(\lambda_{II}(x)+\lambda_{IO}(x))(1-q_I)+\lambda_{IO}(x)pq_I]F(\theta)\\
\nonumber+[\lambda_{OI}(x)(1-p)+\lambda_{OO}(x)(1-p)^2]q_OF(B_O+\theta)\\
+[(\lambda_{OI}(x)+\lambda_{OO}(x))(1-q_O)+(\lambda_{OI}(x)p+\lambda_{OO}(x)(2p-p^2))q_O]F(\theta)
\end{eqnarray}
\\
and the average selection gradient of a mutant can be calculated in the same way as described in Eqs 9-10 of the main text.

Table S2 gives event probabilities for an OR logic decision process, in which behavioral strategies have two components: $p_p^i$ is the probability that a player $i$ is willing to interact with a member of their party and $p_g^i$ is the probability that a player $i$ is willing to interact with a member of their identity group.

\begin{figure}[th!] \centering \includegraphics[scale=0.1]{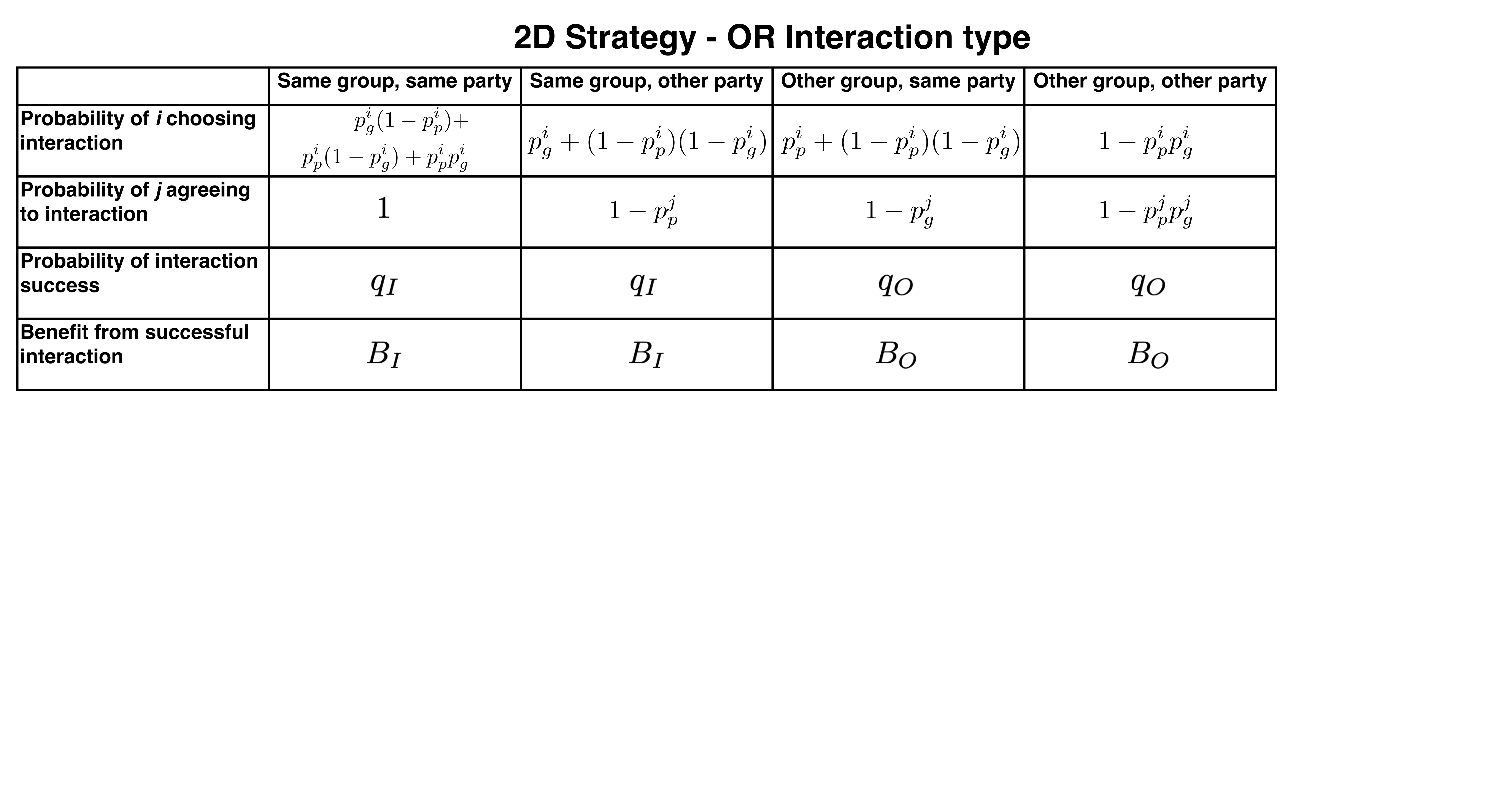}
\caption*{\small Table S2: \textbf{OR decision logic with a two-dimensional strategy}.}
\end{figure}

\noindent The fitness for this process is as described in Eq. 6 and Eq. 8, with the appropriate substitutions from Table S2.

\section*{Adaptive dynamics of polarization}

In this section we present additional results for the ``adaptive dynamics'' analysis of the model. We perform invasion analyses under the assumption that that the population is infinitely large, and all members of the population adopt the same strategy. We then compute the selection gradient experienced by a rare mutant, in order to determine whether it will spread.

\subsection*{Divergent selection pressures}

We first note that Eqs 9-10 of the main text describe the average selection gradient across both groups. However, in general when the distribution of identity group with respect to party is asymmetric, i.e.~$0<|\chi|<1$, a member of group 1 belonging to party 1 will experience different selection pressures than a member of group 1 belonging to party 2, and so on. Under our assumption that identity groups and parties are of equal size and experience the same risk profiles, however, the only equilibria we find either occur when $\chi=0$ (both parties are well mixed with respect to identity group) or $|\chi|=1$ (identity groups align perfectly with parties). Under these conditions selection pressures are symmetric. Since any internal state in which selection pressures are asymmetric is unstable, we are able to ignore the complications that arise in such cases and focus on the stability of the symmetric equilibria. We note however that if different groups experience different risk profiles, or are of different size, this symmetry may not hold and new dynamics may arise.

\subsection*{Only party joint dynamics}

Figure S1 shows the joint dynamics of sorting and polarization under the only party decision process (see main text Table 1, Eq. 5 and Eq. 7), in which individuals make decisions about who to interact with based only on party identity.

\begin{figure}[th!] \centering \includegraphics[scale=0.25]{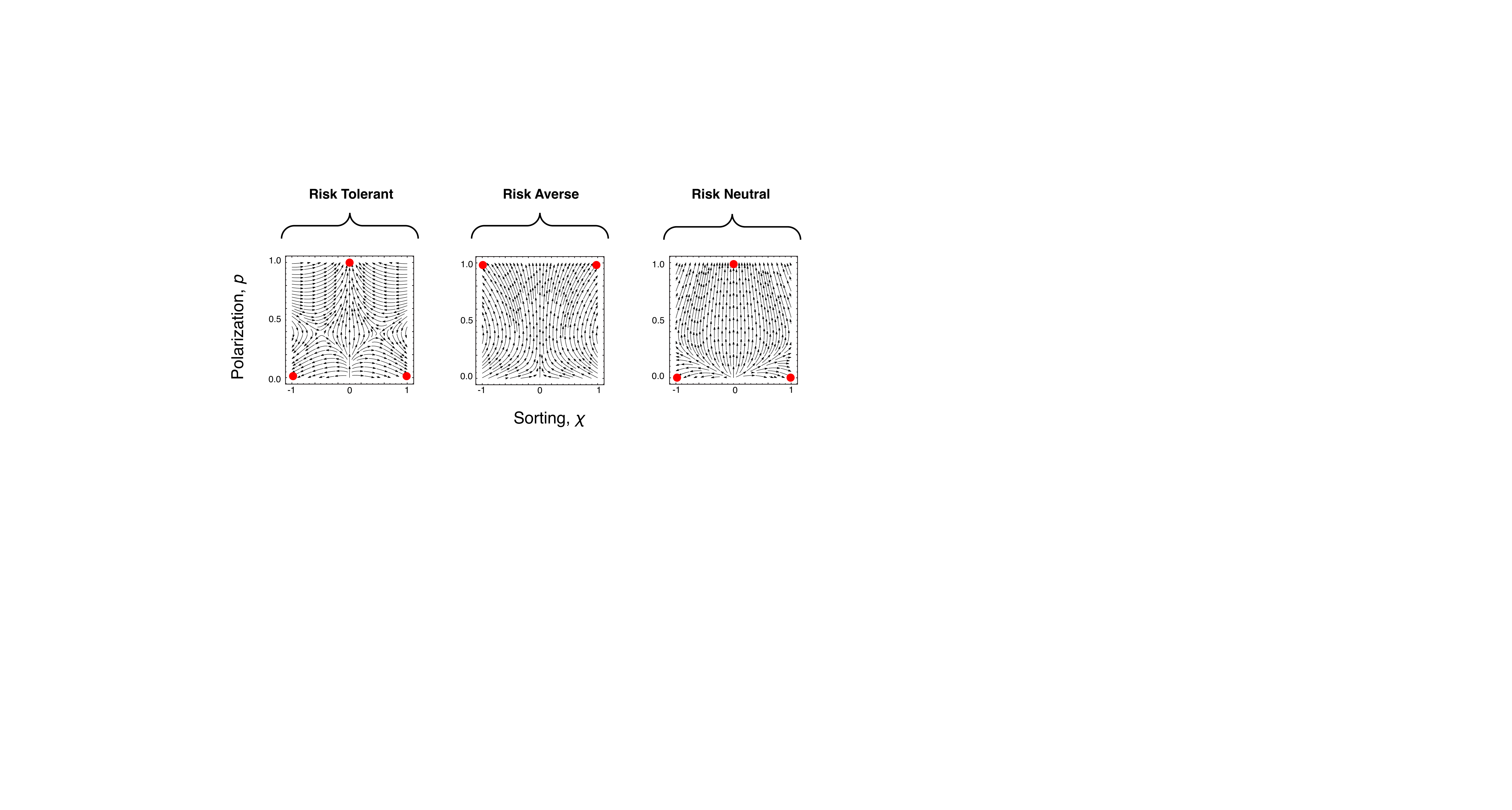}
\caption*{\small Figure S1: \textbf{Only party -- Polarization and sorting}. Phase portraits illustrate the dynamics of polarization $p$ and degree of sorting $\chi$ under our model of economic interactions and party switching, with fixed identity groups. Arrows indicate the average selection gradient experienced by a local mutant against a monomorphic background (see Methods). Red dots indicate stable equilibria. (left) When the decision process for social interactions considers only party identity, both high and low polarization states are stable. When polarization is low, sorting is always high $|\chi|=1$, but when polarization is high, low sorting ($\chi=0$) becomes stable. (center) However when the environment is risk averse, only high polarization and high sorting are stable. (right) And finally when the environment is risk neutral, the system returns to bistable polarization with high sorting when polarization is low, and low sorting when polarization is high. These plots show dynamics for $B_I=0.5$, $B_O=1$, $q_I=1.0$, $q_O=0.6$, $h=10$ and $a=0.02$. The phase portraits here show the selection gradient experienced by a monomorphic population in which parties and groups are of equal size.}
\end{figure}

\clearpage

\subsubsection*{AND logic joint dynamics}

Figure S2 shows the joint dynamics of sorting and polarization under the AND party decision process (see Table S1), in which individuals make decisions about who to interact with based on party and group identity using AND logic.

\begin{figure}[th!] \centering \includegraphics[scale=0.25]{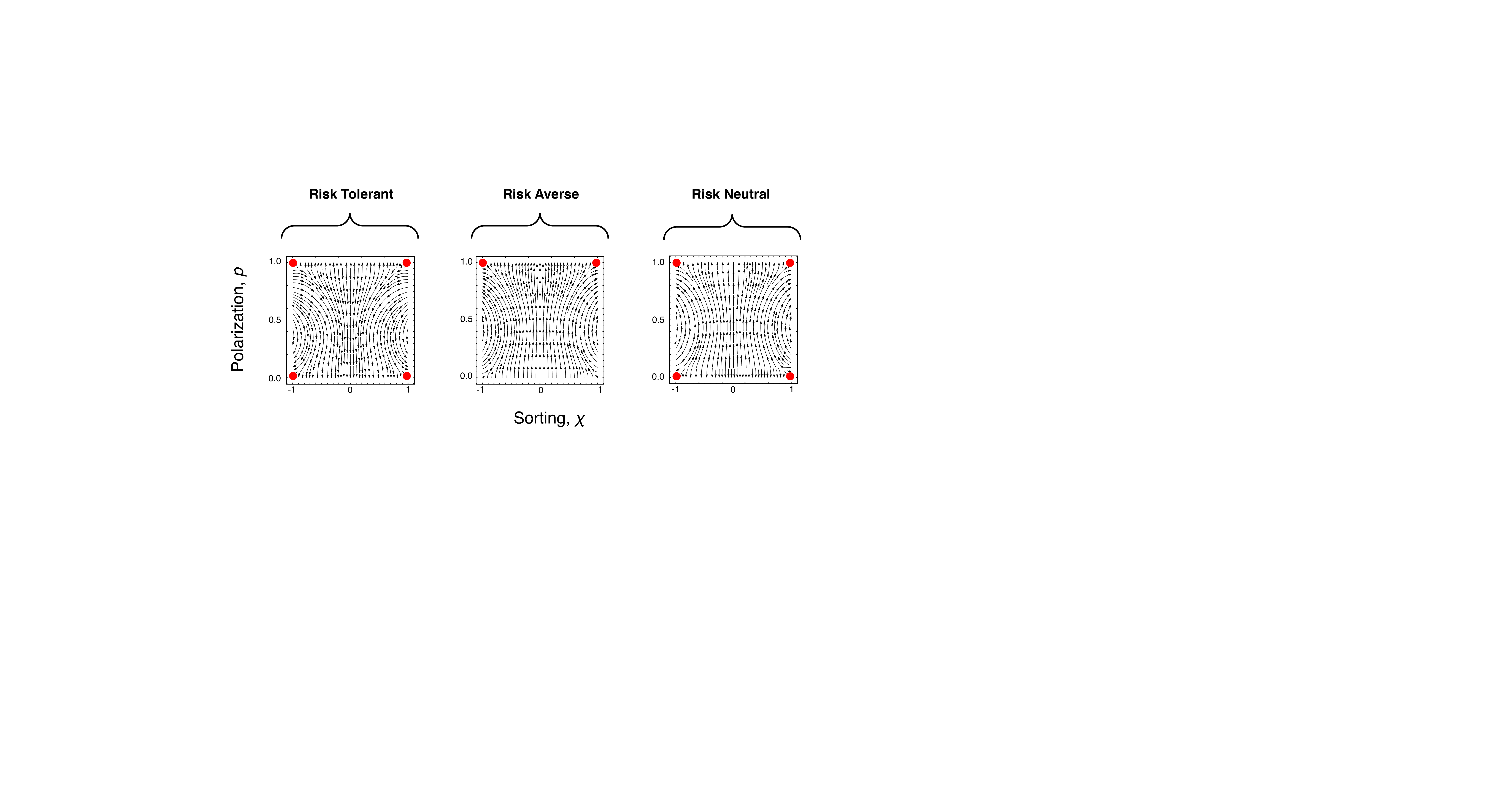}
\caption*{\small Figure S2: \textbf{Party AND group -- Polarization and sorting}. Phase portraits illustrate the dynamics of polarization $p$ and degree of sorting $\chi$ under our model of economic interactions and party switching, with fixed identity groups. Arrows indicate the average selection gradient experienced by a local mutant against a monomorphic background (see Methods). Red dots indicate stable equilibria. (left) When the decision process for social interactions considers group AND party identity, both high and low polarization states are stable, but sorting is always high $|\chi|=1$. (center) However when the environment is risk averse, only high polarization and high sorting are stable. (right) And finally when the environment is risk neutral, the system returns to bistable polarization with high sorting. These plots show dynamics for $B_I=0.5$, $B_O=1$, $q_I=1.0$, $q_O=0.6$, $h=10$ and $a=0.02$. The phase portraits here show the selection gradient experienced by a monomorphic population in which parties and groups are of equal size.}
\end{figure}

\clearpage

\subsection*{Joint dynamics of group and party polarization}

Figure S3 shows the joint dynamics of group and identity polarization under the OR party decision process (see  Table S2), in which individuals make decisions about who to interact with based on party and group identity using OR logic, and strategies are two-dimensional, meaning that an individual can weight the two dimensions of identity differently. We see that across all environments, there are two equilibria, with either high group and low party polarization, or vice versa.

\begin{figure}[th!] \centering \includegraphics[scale=0.25]{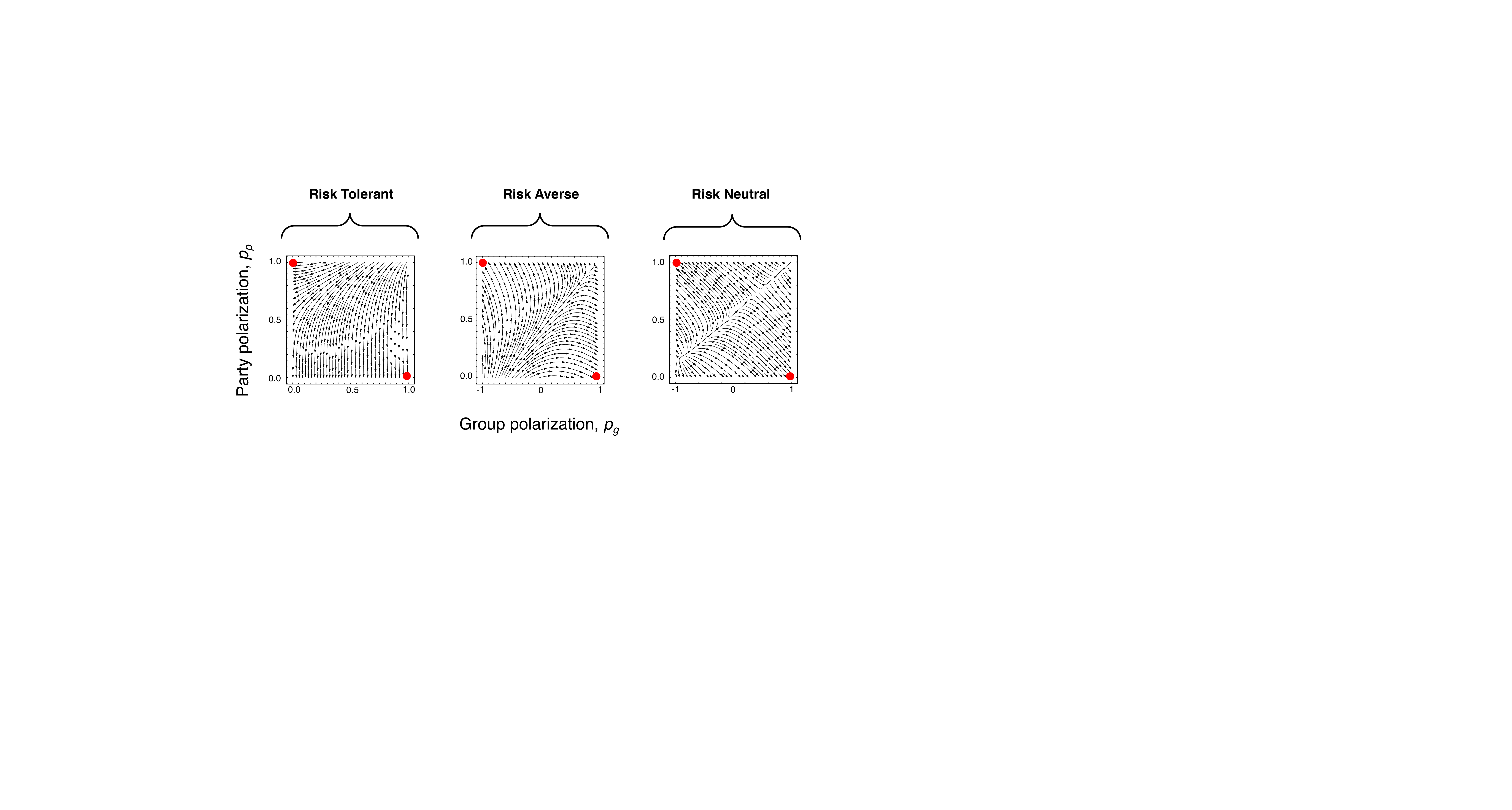}
\caption*{\small Figure S3: \textbf{Party OR group -- Group and identity polarization}. Phase portraits illustrate the dynamics of group polarization $p_g$ and party polarization $p_p$ under our model of economic interactions, with high sorting $\chi=1$. Arrows indicate the average selection gradient experienced by a local mutant against a monomorphic background (see Methods). Red dots indicate stable equilibria. When the decision process for social interactions considers group and party identity using OR logic, either high group polarization $p_g=1$ and low polarization $p_p=0$ or $p_p=1$ and low polarization $p_g=0$ are stable. These plots show dynamics for $B_I=0.5$, $B_O=1$, $q_I=1.0$, $q_O=0.6$, $h=10$ and $a=0.02$. The phase portraits here show the selection gradient experienced by a monomorphic population in which parties and groups are of equal size.}
\end{figure}

\clearpage

\subsubsection*{Bistability}

We now describe the evolution of polarization under fixed levels of sorting, $\chi$, for each of the three different decision processes described in Table 1 of the main text, across different economic environments, $\theta$ (Figure S4). We see that high polarization strategies are the only stable outcome in risk-adverse economic environments ($\theta\approx 0$), whereas the availability of stable low polarization strategies, and the associated basin of attraction for such strategies, varies with the economic environment, the decision process and the degree of sorting in the population.

\begin{figure}[th!] \centering \includegraphics[scale=0.11]{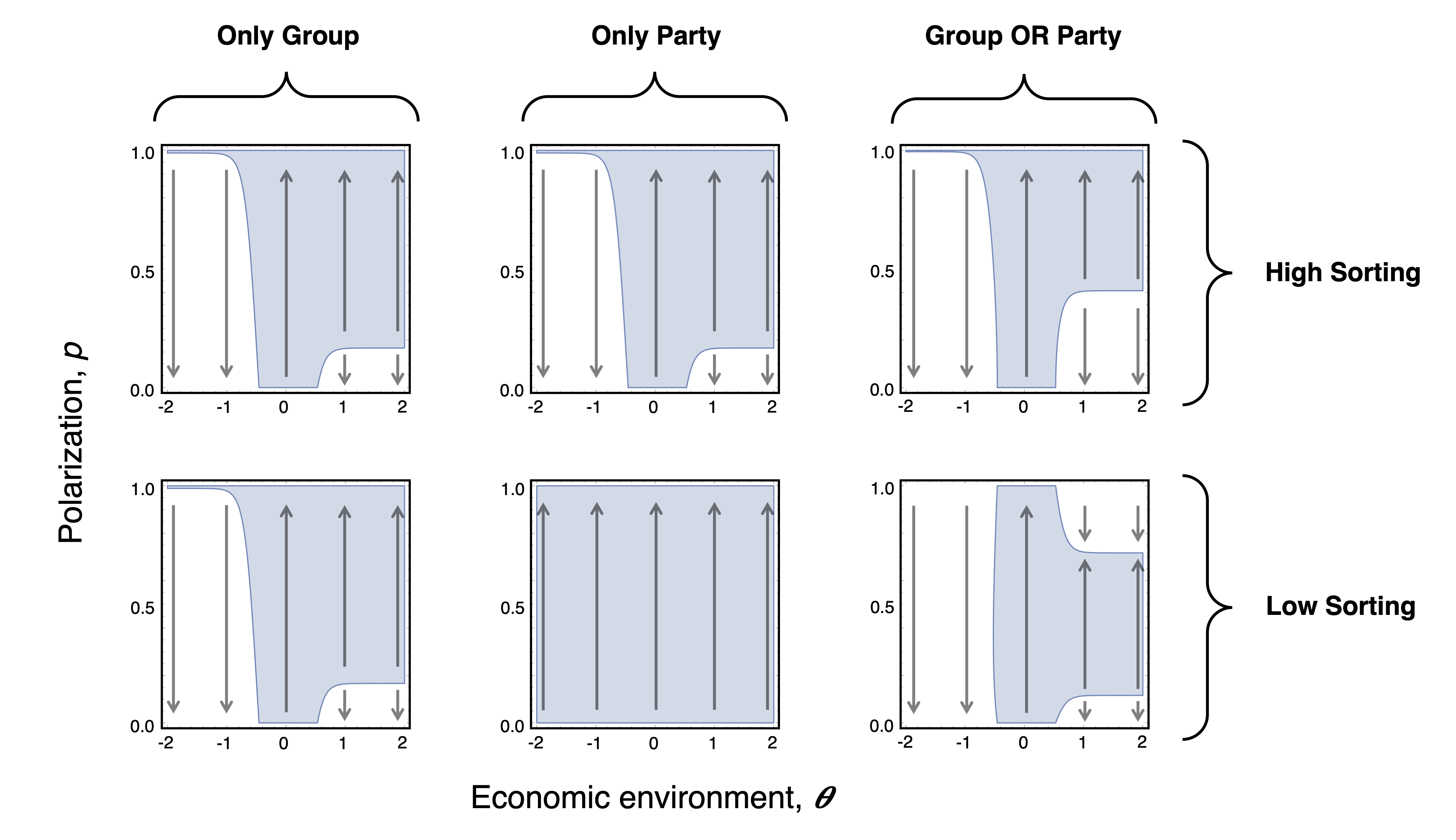}
\caption*{\small Figure S4 \textbf{Polarization across economic environments}. Shown are the evolutionary dynamics of polarization $p$, i.e. the probability of choosing in-group or in-party interactions, for different underlying economic environments $\theta$. Arrows indicate the selection gradient (see Methods) experienced by a monomorphic population employing strategy $p$ in environment $\theta$. Blue regions indicate a positive gradient (increasing polarization) while white regions indicate a negative gradient (declining polarization). Six cases are presented. (top, left) When only group identity is attended to, polarization is bistable provided the environment is not risk averse. However under risk aversion ($\theta\approx 0$), only high polarization is stable. (bottom, left) The dynamics do not depend on sorting when only group identity is attended to. (top, center) When only party identity is attended to and sorting is high ($|\chi|=1$), the dynamics and basins of attraction for group and party identity are identical. (bottom center) However when sorting is low ($\chi=0$), high polarization always evolves in a population only attends to party identity. (top, right) When interaction decisions are based on party or group identity, and sorting is high, the basin of attraction for low polarization is large compared to the case when decisions are based on party or group alone, meaning that polarization is comparatively stable. (bottom, right) When sorting is low, low polarization becomes the only stable equilibrium under risk tolerance, while an intermediate polarization state becomes stable under risk aversion. These plots show dynamics for $B_I=0.5$, $B_O=1$, $q_I=1.0$, $q_O=0.6$, $h=10$ and $a=0.02$. Dynamics describe the selection gradient experienced by a monomorphic population in which parties and groups are of equal size.}
\end{figure}

Most strikingly, when only party identity is used to make decisions, low levels of sorting result in high polarization regardless of the economic environment; whereas a strategy that uses party \emph{or} group identity to make decisions tends to increase the basin of attraction to low-polarization strategies. In risk-tolerant environments ($\theta<0$), the low-polarization outcome may even be the only stable strategy.

The intuition for this is simple: when sorting is low, a decision process that only accounts for party identity provides little information about the likely success of interactions. In contrast, a decision process that uses party identity or group identity to define the in-group widens the pool of potential out-group interaction partners, and the associated increased benefits of those interactions. However such a widening of the pool is of no use when the environment favors risk aversion ($\theta\approx 0$, Figure S4). 

\clearpage

\subsection*{Evolution of attention}

We now consider whether a population can be incentivized to switch from a group-only decision process to a group- or party-decision process (see Table 1 of the main text and Table S2). We calculated the fitness of each decision process as a function of sorting $\chi$ and group polarization strategy $p_g$. We assume that initially $p_p=0.5$ for the OR logic decision strategy, indicating indifference to party identity.

\begin{figure}[th!] \centering \includegraphics[scale=0.25]{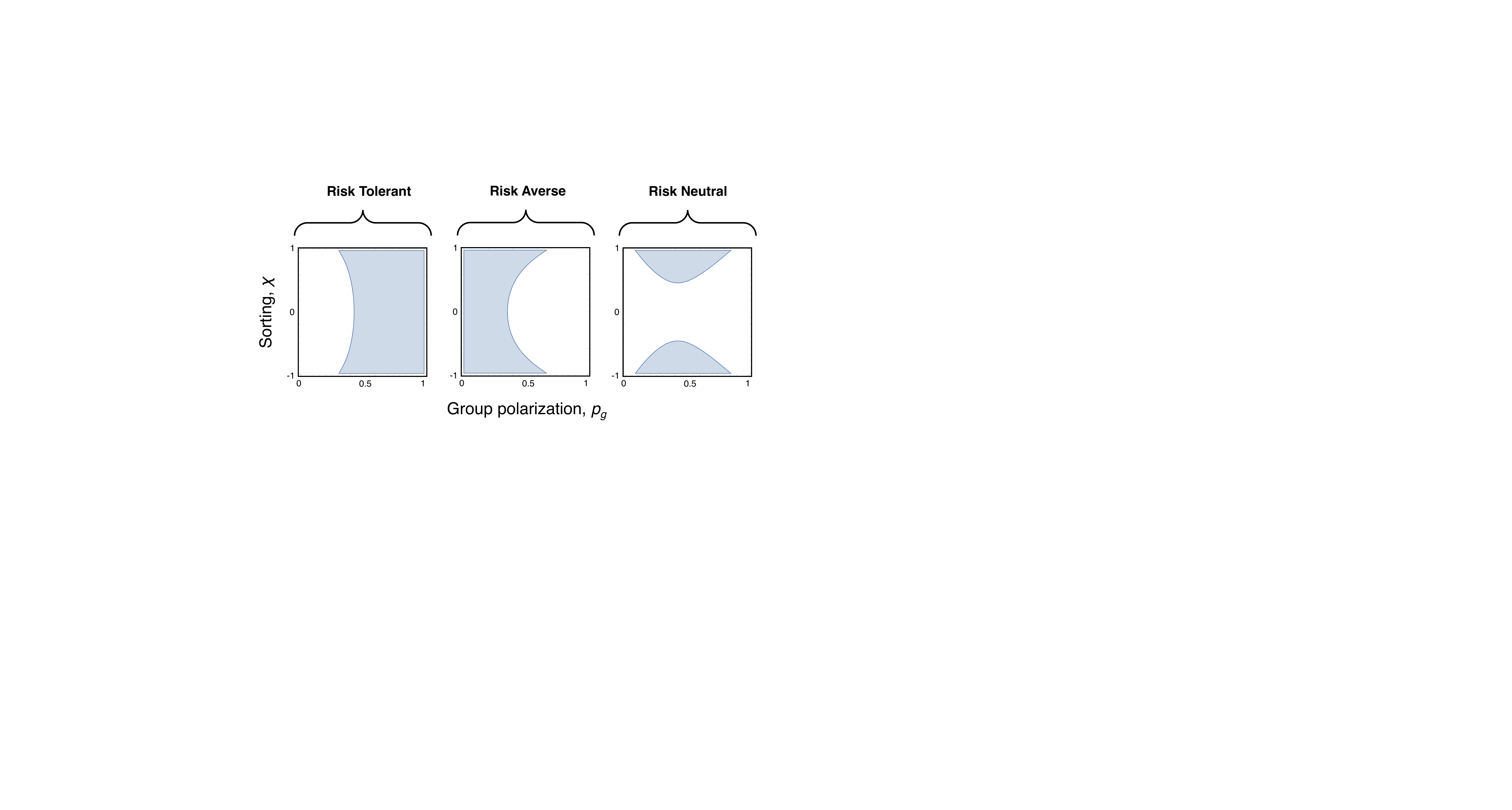}
\caption*{\small Figure S5: \textbf{Switching decision logic}. Shown in blue are the conditions under which an OR decision logic (Table S2) produces higher fitness than a group only decision logic, as a function of sorting, $\chi$ and group polarization strategy $p_g$ in a monomorphic population. We assume that the OR decision logic uses $p_p=0.5$ indicating initial indifference to party identity (i.e neither in-party nor out-party favoring). (left) When the environment is risk tolerant, high polarization group only decision strategies can be invaded by group OR polarization decision strategies. (center) When the environment is risk averse, low polarization group only decision strategies can be invaded by group OR polarization decision strategies. (right) When the environment is risk neutral, highly sorted populations with intermediate levels of polarization under a group only decision strategy can be invaded by group OR polarization decision strategies. These plots show dynamics for $B_I=0.5$, $B_O=1$, $q_I=1.0$, $q_O=0.6$, $h=10$ and $a=0.02$. Dynamics describe the selection gradient experienced by a monomorphic population in which parties and groups are of equal size.}
\end{figure}

\clearpage

\subsection*{Varying model parameters}

Here we vary the parameters for the payoffs and for the utility function, under the group OR party decision logic. We show that the bi-stability of the system is highly robust to parameter variation, but if the utility function becomes too shallow, risk aversion does not stimulate polarization.

\subsubsection*{Varying payoffs}

We varied the value of $B_O$, keeping other parameters fixed, and retaining the risk profile in which in-group interactions are less risky but less advantageous.

\begin{figure}[th!] \centering \includegraphics[scale=0.25]{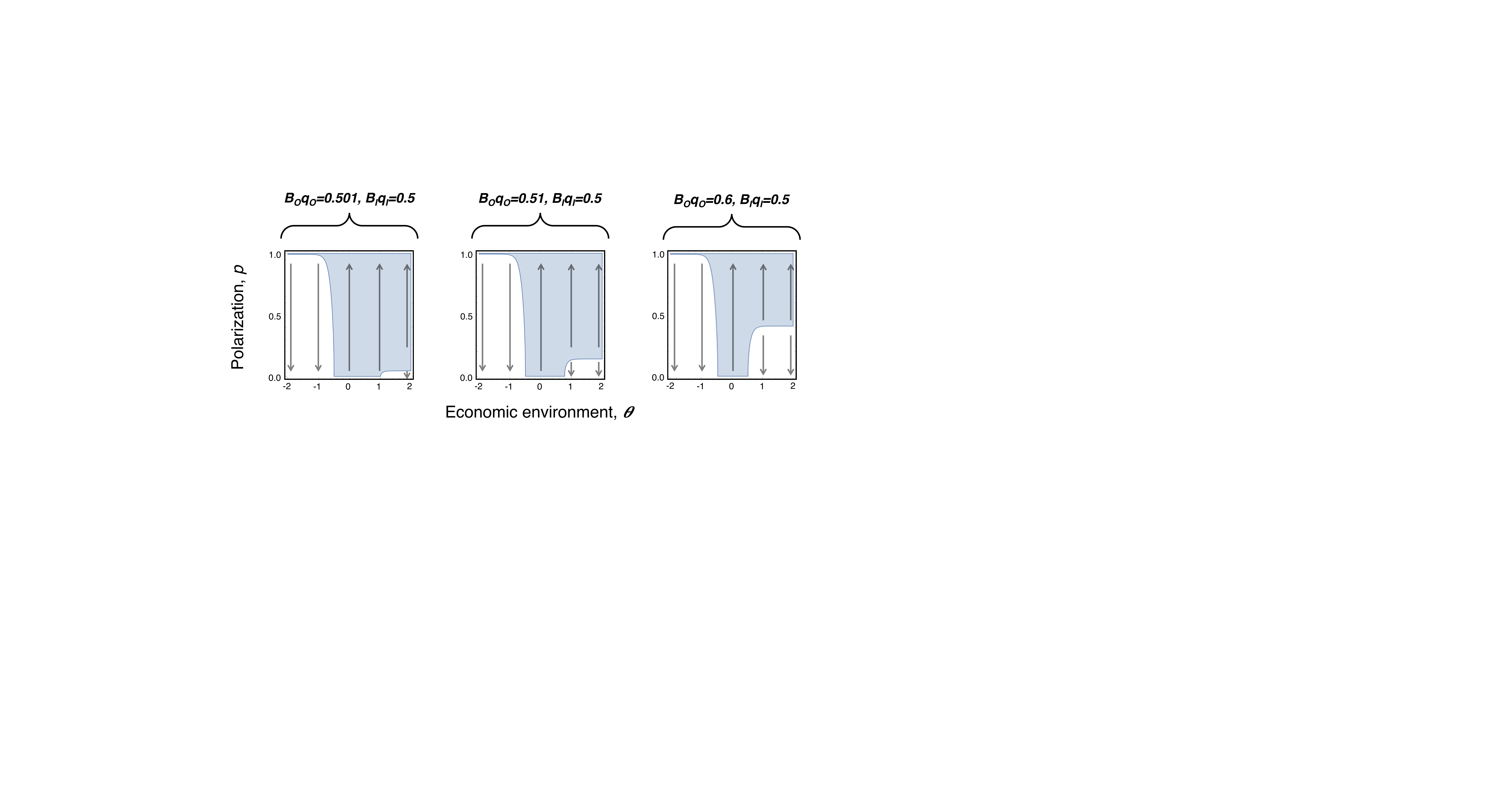}
\caption*{\small Figure S6: \textbf{Polarization under different payoffs}. Shown are the evolutionary dynamics of polarization $p$, i.e. the probability of choosing in-group or in-party interactions, for different underlying economic environments $\theta$. Arrows indicate the selection gradient (see Methods) experienced by a monomorphic population employing strategy $p$ in environment $\theta$. Blue regions indicate a positive gradient (increasing polarization) while white regions indicate a negative gradient (declining polarization). In all cases we show the group OR party decision logic as described in main text Table 1. (left) When the expected benefit of out-group interactions is only slightly advantageous ($B_oq_o=0.501$, $B_iq_i=0.5$), polarization is bistable provided the environment is not risk averse. Under risk aversion ($\theta\approx 0$), only high polarization is stable. However when the environment is risk neutral, the basin of attraction for the low polarization equilibrium is small. (center) When the expected benefit of out-group interactions is moderately advantageous ($B_oq_o=0.51$, $B_iq_i=0.5$), polarization is bistable provided the environment is not risk averse. Under risk aversion ($\theta\approx 0$), only high polarization is stable. When the environment is risk neutral, the basin of attraction for the low polarization increases. (right) When the expected benefit of out-group interactions is more advantageous ($B_oq_o=0.6$, $B_iq_i=0.5$), polarization is bistable provided the environment is not risk averse. Under risk aversion ($\theta\approx 0$), only high polarization is stable. When the environment is risk neutral, the basin of attraction for the low polarization increases further. These plots show dynamics for $B_I=0.5$, $B_O=1$, $q_I=1.0$, $q_O=0.6$, $h=10$ and $a=0.02$. Dynamics describe the selection gradient experienced by a monomorphic population in which parties and groups are of equal size.}
\end{figure}

\clearpage

\subsubsection*{Varying utility function non-linear component}

We varied the sharpness of the non-linear component of the utility function, $h$. We see that risk aversion is not sufficient to ensure the evolution of polarization when the $h$ is small and the slope of the S-curve is shallow.

\begin{figure}[th!] \centering \includegraphics[scale=0.25]{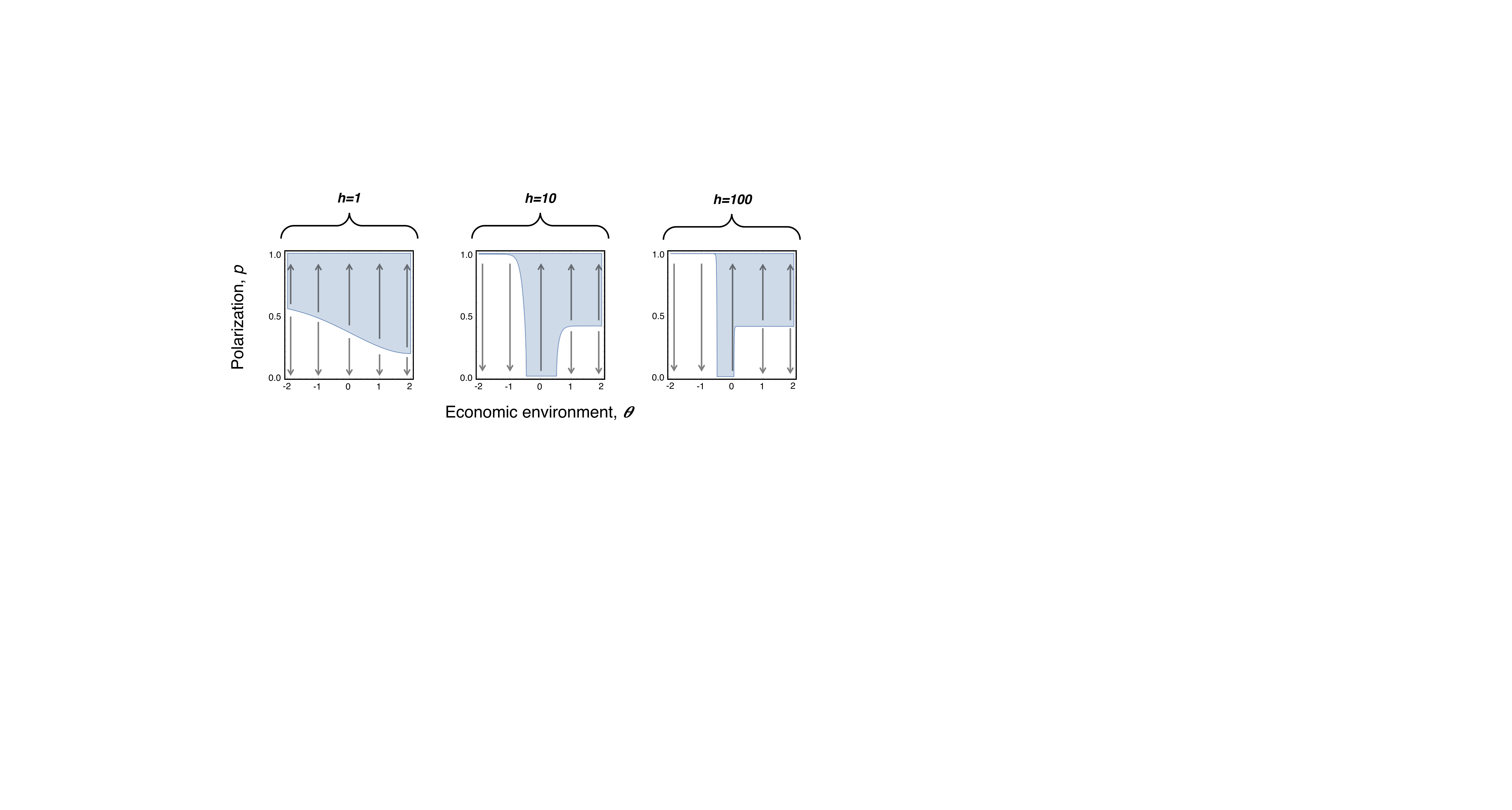}
\caption*{\small Figure S7: \textbf{Polarization under different utility function sharpness, $h$}. Shown are the evolutionary dynamics of polarization $p$, i.e. the probability of choosing in-group or in-party interactions, for different underlying economic environments $\theta$. Arrows indicate the selection gradient (see Methods) experienced by a monomorphic population employing strategy $p$ in environment $\theta$. Blue regions indicate a positive gradient (increasing polarization) while white regions indicate a negative gradient (declining polarization). In all cases we show the group OR party decision logic as described in main text Table 1. (left) When the sharpness of the utility function threshold is shallow, $h=1$, polarization is always bistable. (center) When the sharpness of the utility function threshold is intermediate, $h=10$, polarization is bistable provided the environment is not risk averse. Under risk aversion ($\theta\approx 0$), only high polarization is stable. (right) When the sharpness of the utility function threshold is sharp, $h=100$, polarization is bistable provided the environment is not risk averse. Under risk aversion ($\theta\approx 0$), only high polarization is stable. These plots show dynamics for $B_I=0.5$, $B_O=1$, $q_I=1.0$, $q_O=0.6$, $h=10$ and $a=0.02$. Dynamics describe the selection gradient experienced by a monomorphic population in which parties and groups are of equal size.}
\end{figure}

\clearpage

\subsubsection*{Varying utility function linear component}

We varied the slope of the linear component of the utility function, $a$. We see that the qualitative dynamics of the system are robust to the choice of $a$ over two orders of magnitude.

\begin{figure}[th!] \centering \includegraphics[scale=0.25]{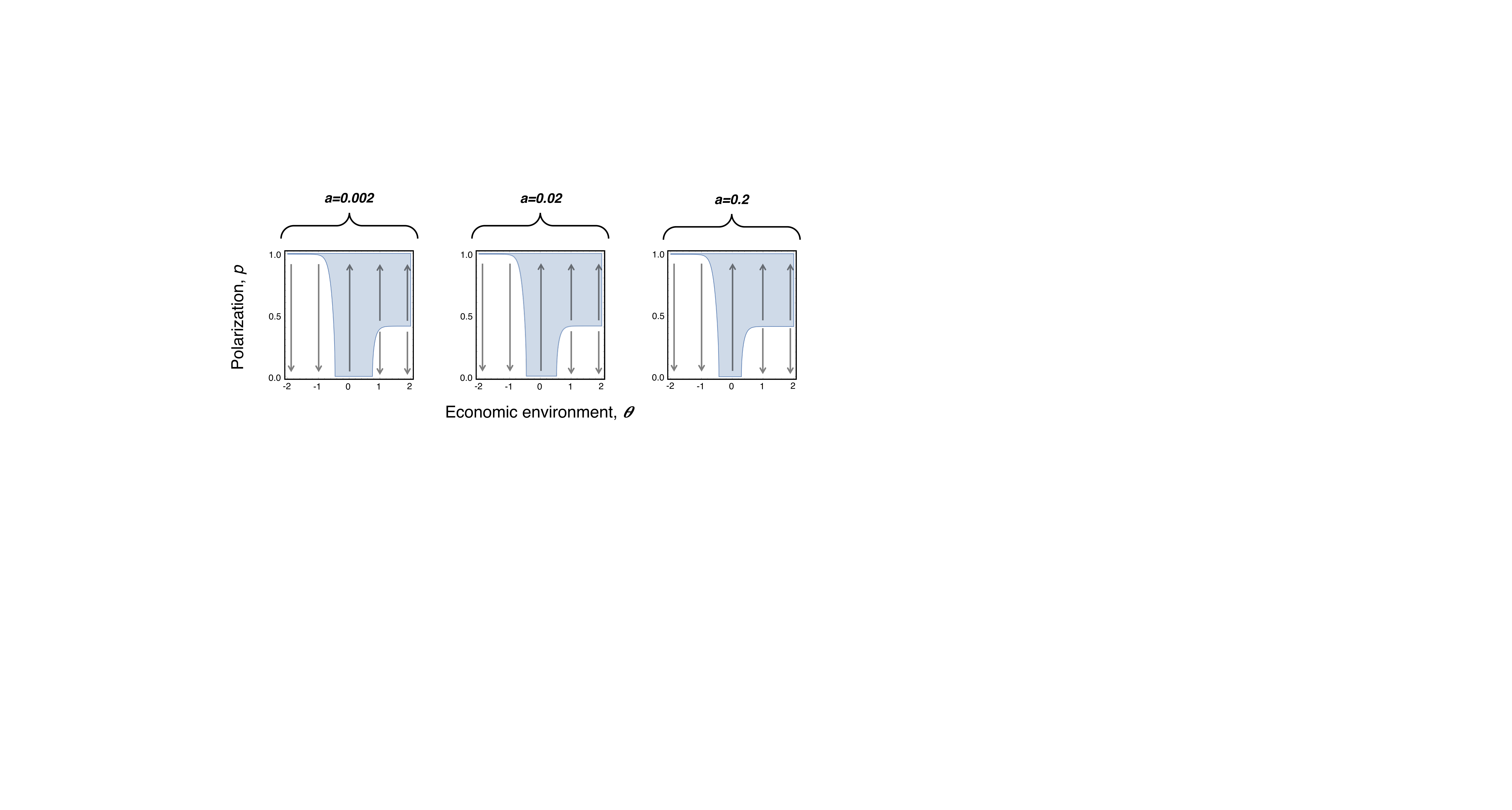}
\caption*{\small Figure S8: \textbf{Polarization under different utility function linear component slope, $a$}. Shown are the evolutionary dynamics of polarization $p$, i.e. the probability of choosing in-group or in-party interactions, for different underlying economic environments $\theta$. Arrows indicate the selection gradient (see Methods) experienced by a monomorphic population employing strategy $p$ in environment $\theta$. Blue regions indicate a positive gradient (increasing polarization) while white regions indicate a negative gradient (declining polarization). In all cases we show the group OR party decision logic as described in main text Table 1. Regardless of the the slope of the linear part of utility function, $a$, polarization is bistable provided the environment is not risk averse. Under risk aversion ($\theta\approx 0$), only high polarization is stable. These plots show dynamics for $B_I=0.5$, $B_O=1$, $q_I=1.0$, $q_O=0.6$, $h=10$ and $a=0.02$. Dynamics describe the selection gradient experienced by a monomorphic population in which parties and groups are of equal size.}
\end{figure}

\clearpage

\section*{Simulations of heterogeneous populations}

We now discuss additional simulations, conducted under the same assumptions as described in the main text.

\subsection*{Sorting}

We exogenously varied the amount of sorting, $\chi$. We see that sorting tends to increase polarization, but it can have complex effects on levels of inequality and population average utility. This is because intermediate levels of polarization tend to result in lower levels of utility. Where reducing sorting can also reduce polarization to low levels, it has a beneficial effect in reducing inequality and increasing population average utility.

\begin{figure}[th!] \centering \includegraphics[scale=0.2]{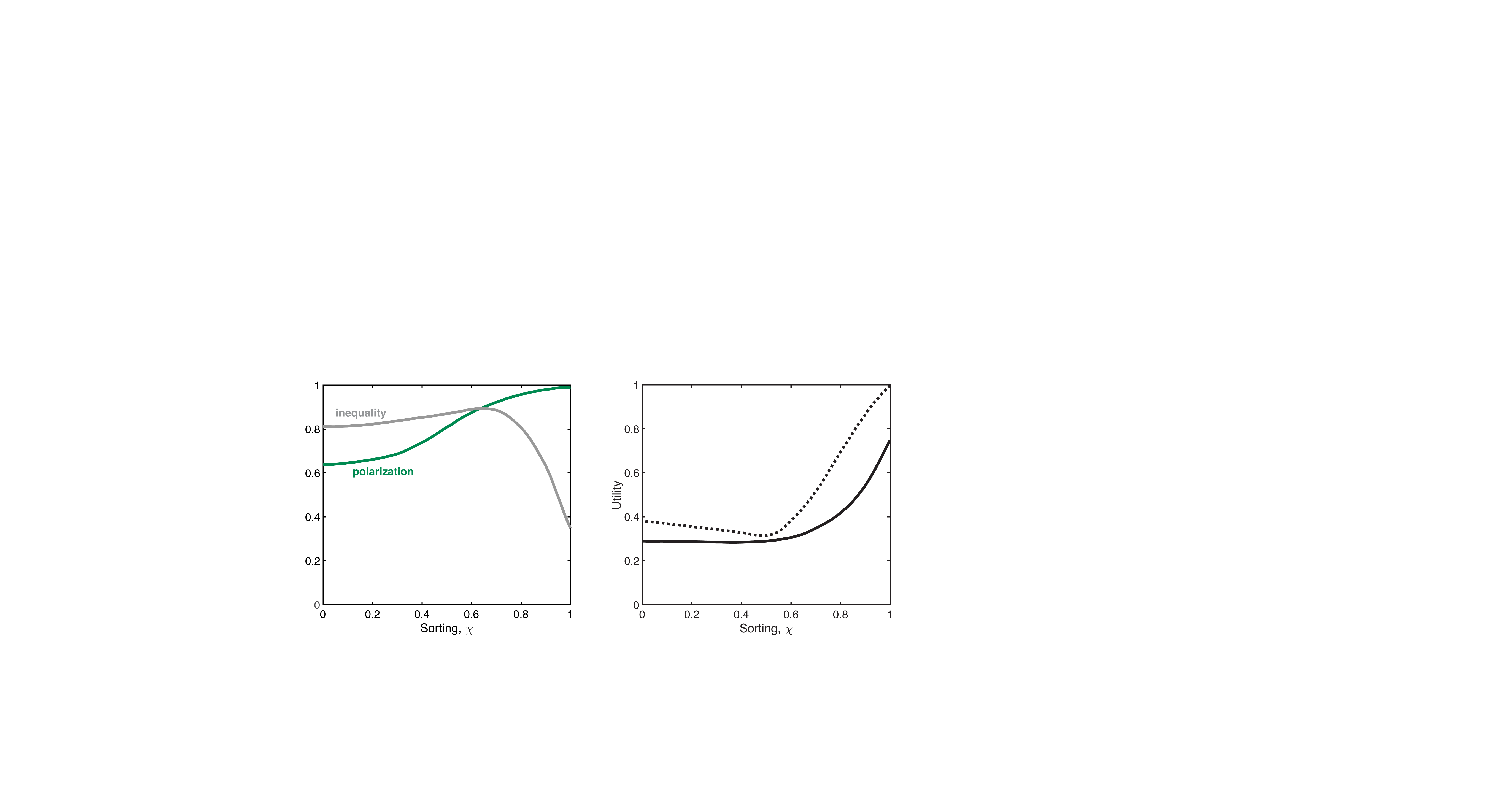}
\caption*{Figure S9: \small \textbf{Sorting and inequality.} Ensemble mean equillibria from individual-based simulations  for a population initialized in a low polarization state in the presence of wealth redistribution (Eq.~4). We show results in the case no underlying economic inequality, $\beta=0.5$ (dashed lines), as well the case of high underlying inequality, $\beta=0.01$ (solid lines). Results shown here arise from a decision process that attends to group or party identity, and redistribution is fixed exogenously at $\alpha=1$. (a) When public goods are not multiplicative ($r=1$ and $\theta_0=0.5$), as sorting increases, polarization increases, but inequality changes non-linearly, achieving its lowest value at maximum sorting. b) Increasing sorting tends to increases overall utility. Plots show ensemble mean values across $10^4$ replicate simulations, for groups of 1000 individuals each. Success probabilities and benefits are fixed at $B_I=0.5$, $B_O=1$, $q_I=1.0$, $q_O=0.6$ with $h=10$ and $a=0.02$, while $\gamma=0$. Evolution occurs via the copying process (see methods) with selection strength $\sigma=10$, mutation rate $\mu=10^{-3}$ and mutation size $\Delta=0.01$.}
\end{figure}

\clearpage

\subsection*{Public goods multiplication factor}

We explored the effect of redistribution on inequality when public goods have a multiplicative effect, $r>1$.

\begin{figure}[th!] \centering \includegraphics[scale=0.2]{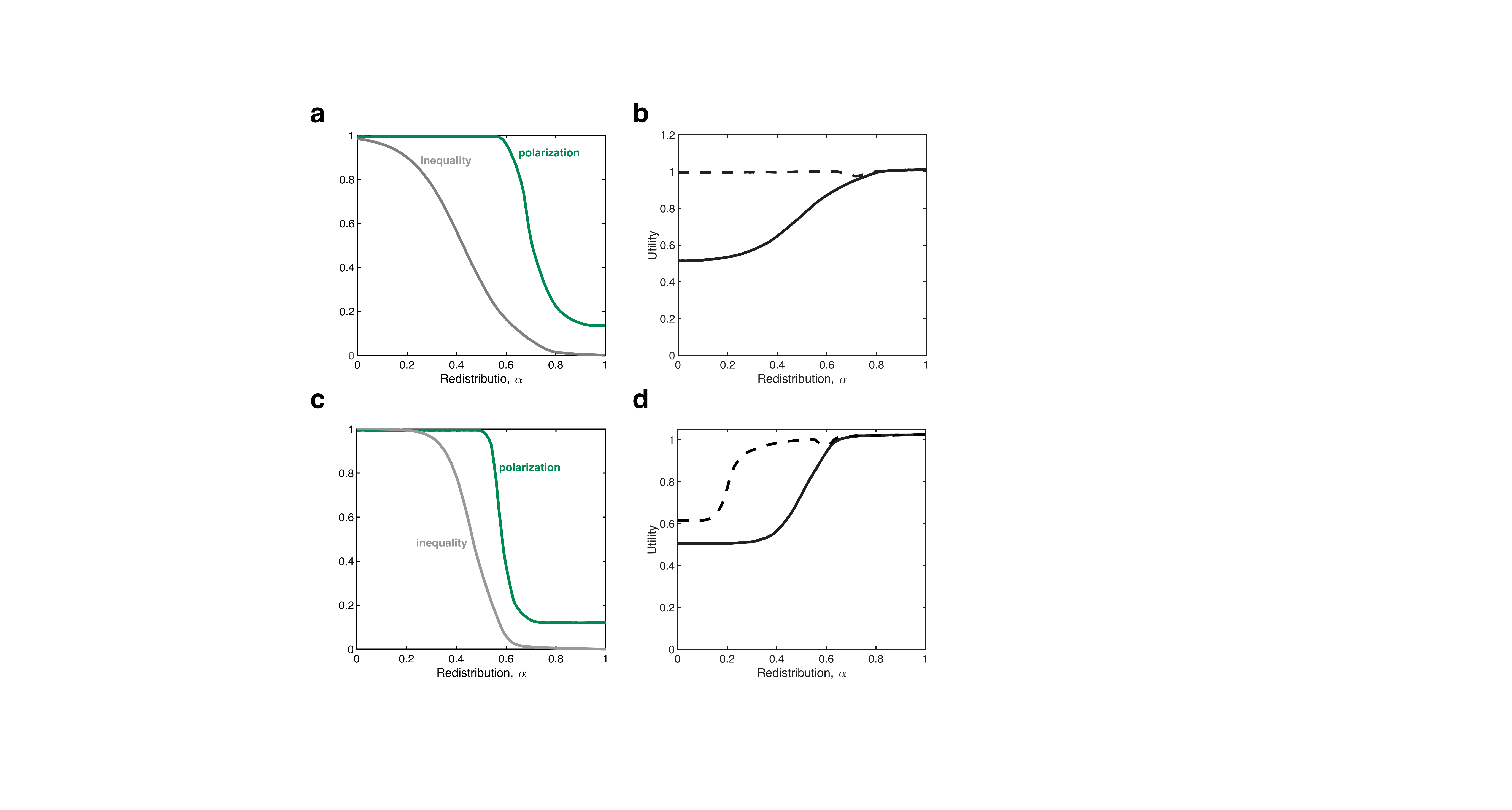}
\caption*{\small Figure S10: \small \textbf{Multiplicative public goods.} Ensemble mean equilibria from individual-based simulations  for a population initialized in a low polarization state in the presence of wealth redistribution (Eq.~4). We show results in the case no underlying economic inequality, $\beta=0.5$ (dashed lines), as well the case of high underlying inequality, $\beta=0.01$ (solid lines). Results shown here arise from a decision process that attends to group or party identity, and sorting is fixed exogenously at $\chi=1$. When public goods are not multiplicative ($r=1$ and $\theta_0=0.5$), and redistribution is absent ($\alpha=0$) overall inequality (gray line, measured as the relative difference in utility -- see SI) and polarization (green line) are high. With increasing rates of redistribution, first overall inequality and then polarization decline to zero. b) Increasing redistribution increases overall utility towards the level achieved when underlying inequality is absent. b) When public goods are multiplicative ($r=2$ and $\theta_0=1$), we see the same dynamics, but lower levels of redistribution are required to reduce inequality and polarization. d) In this case increasing redistribution also increases the utility of the population when inequality is absent (dashed line). Plots show ensemble mean values across $10^4$ replicate simulations, for groups of 1000 individuals each. Success probabilities and benefits are fixed at $B_I=0.5$, $B_O=1$, $q_I=1.0$, $q_O=0.6$ with $h=10$ and $a=0.02$, while $\gamma=0$. Evolution occurs via the copying process (see methods) with selection strength $\sigma=10$, mutation rate $\mu=10^{-3}$ and mutation size $\Delta=0.01$.}
\end{figure}

\clearpage

\subsection*{Loss due to taxation}

We explored the effect of deadweight losses due to taxation, $\gamma>0$, on the effectiveness of redistribution on reducing polarization and inequality. We see that such losses reduce the effectiveness of redistribution and make it harder to mitigate both polarization and inequality.

\begin{figure}[th!] \centering \includegraphics[scale=0.15]{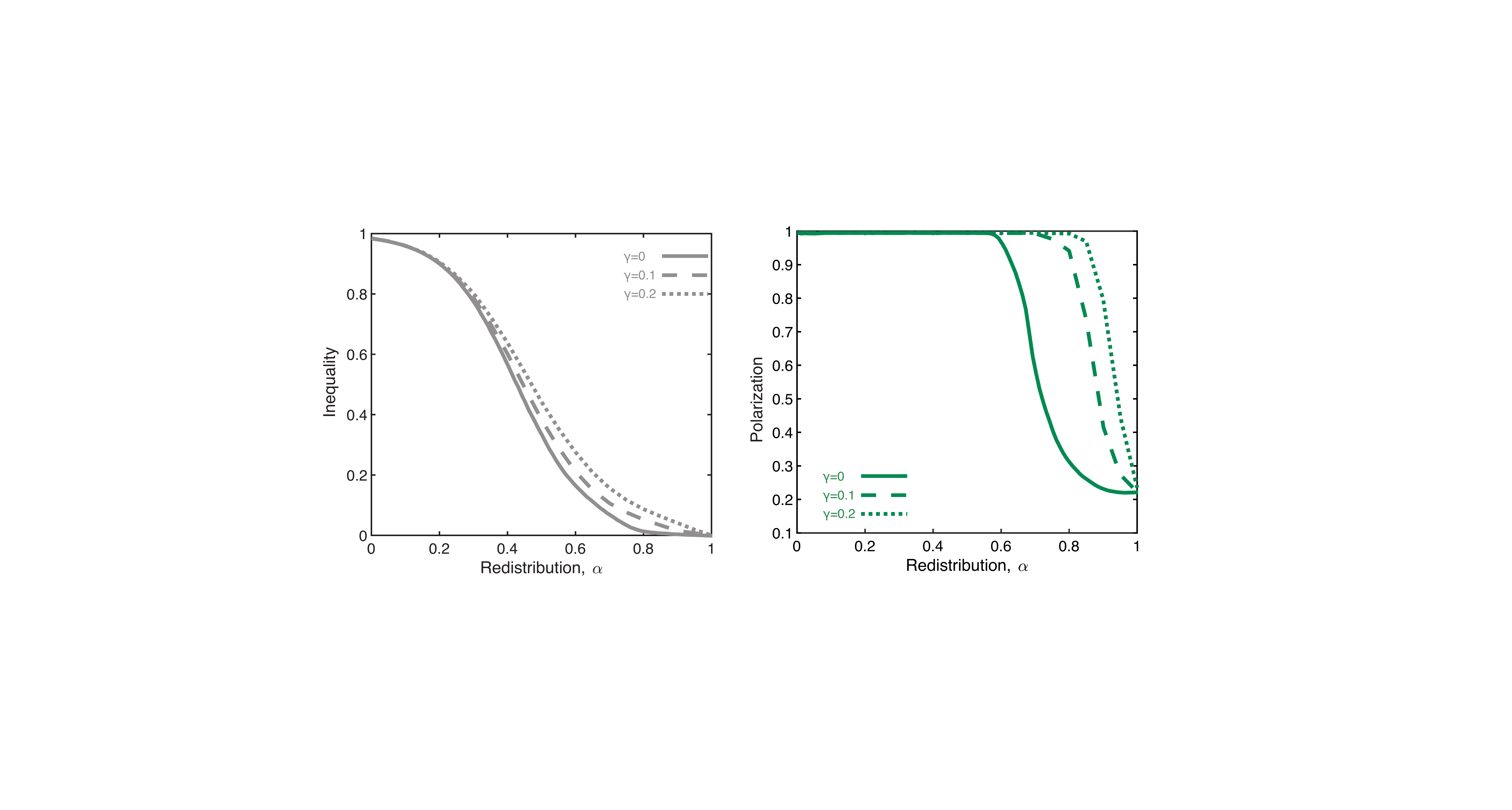}
\caption*{\small Figure S11: \small \textbf{Loss due to taxation.} Ensemble mean equillibria from individual-based simulations for a population initialized in a low polarization state in the presence of wealth redistribution (Eq.~4). We show results for the case of no underlying economic inequality, $\beta=0.5$ (dashed lines), as well the case of high underlying inequality, $\beta=0.01$ (solid lines). Results shown here arise from a decision process that attends to group or party identity, and sorting is fixed exogenously at $\chi=1$. (left) When public goods are not multiplicative ($r=1$ and $\theta_0=0.5$), and redistribution is absent ($\alpha=0$) overall inequality is high. With increasing rates of redistribution, polarization declines, but the decline is slower when deadweight losses due to taxation increase.  (right) Increasing redistribution increases overall utility towards the level achieved when underlying inequality is absent. b) Similarly, redistribution decreases polarization, but becomes less effective as $\gamma$ increases. Plots show ensemble mean values across $10^4$ replicate simulations, for groups of 1000 individuals each. Success probabilities and benefits are fixed at $B_I=0.5$, $B_O=1$, $q_I=1.0$, $q_O=0.6$ with $h=10$ and $a=0.02$. Evolution occurs via the copying process (see methods) with selection strength $\sigma=10$, mutation rate $\mu=10^{-3}$ and mutation size $\Delta=0.01$.}\end{figure}

\clearpage

\subsection*{Economic shocks}

As shown above (Figure S4) the basin of attraction for the high polarization equilibrium declines to the point of almost vanishing in a risk tolerant environment. Therefore we considered a scenario in which a high polarization population enters a very poor economic environment due to an economic shock. We see that, when this occurs, and we fix $\theta=-1.5$, a population will evolve from a state of high to a state of low polarization, whereas it will remain in a state of high polarization in a risk averse, $\theta=0$ or risk neutral $\theta=1.5$ environment.

\begin{figure}[th!] \centering \includegraphics[scale=0.125]{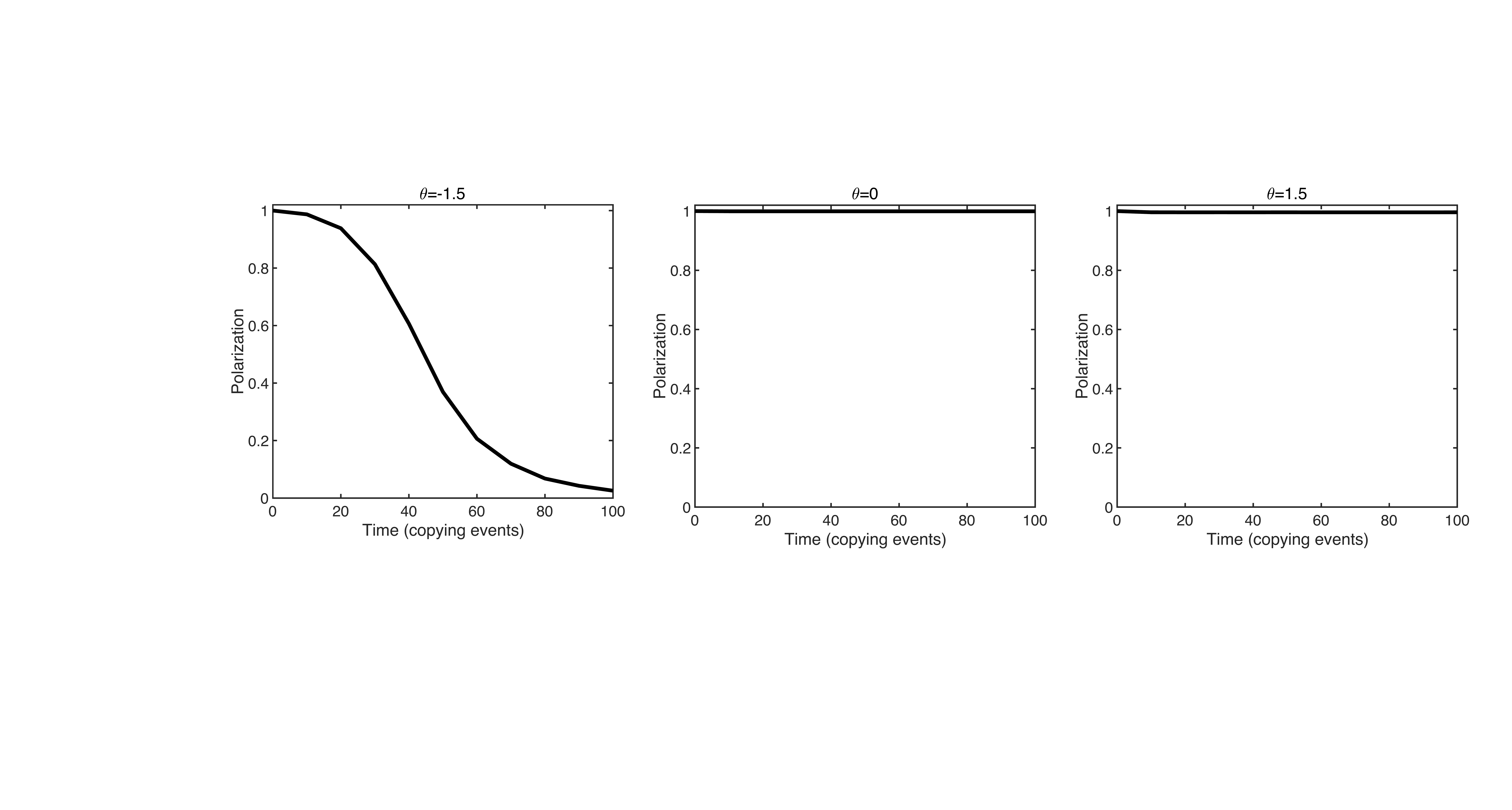}
\caption*{\small Figure S12: \small \textbf{Economic shocks.} Ensemble mean time trajectories for a population initialized in a high polarization state, from individual-based simulations. We show results in the case no underlying economic inequality, $\beta=0.5$ for different (fixed) environments ($\theta$). Results shown here arise from a decision process that attends to group or party identity, with sorting is fixed exogenously at $\chi=1$ and redistribution at $\alpha=0$. (left) When the environment is very bad, $\theta=-0.5$, corresponding to an economic shock, the basin of attraction for the high polarization equilibrium shrinks sufficiently that even small mutations are enough to escape. (center and right) When the environment is not risk tolerant, the population remains stuck in the high polarization equilibrium. Plots show ensemble mean values across $10^4$ replicate simulations, for groups of 1000 individuals each. Success probabilities and benefits are fixed at $B_I=0.5$, $B_O=1$, $q_I=1.0$, $q_O=0.6$ with $h=10$ and $a=0.02$. Evolution occurs via the copying process (see methods) with selection strength $\sigma=10$, mutation rate $\mu=10^{-3}$ and mutation size $\Delta=0.01$.}
\end{figure}

\end{document}